\documentclass[aps,pre,twocolumn,groupedaddress,longbibliography]{revtex4-1}
\usepackage{graphicx,amsmath,amssymb,color,array}
\begin{document}

\title{Densest versus jammed packings of bent-core trimers}
\author{Austin D. Griffith}
\author{Robert S. Hoy}
\email{rshoy@usf.edu}
\affiliation{Department of Physics, University of South Florida}
\date{\today}
\begin{abstract}
We identify putatively maximally dense packings of tangent-sphere trimers with fixed bond angles ($\theta = \theta_0$) using a novel method, and contrast them to the disordered jammed states they form under quasistatic and dynamic  athermal compression.  
Incommensurability of $\theta_0$ with 3D close-packing does not by itself inhibit formation of dense 3D crystals; all $\theta_0$ allow formation of crystals with $\phi_{max}(\theta_0) > 0.97\phi_{cp}$.  
Trimers are always able to arrange into periodic structures composed of close-packed bilayers or trilayers of triangular-lattice planes, separated by ``gap layers'' that accomodate the incommensurability.
All systems have $\phi_J$ significantly below the monomeric value, indicating that trimers' quenched bond-length and bond-angle constraints always act to promote jamming.
$\phi_J$ varies strongly with $\theta_0$; straight ($\theta_0 = 0$) trimers minimize $\phi_J$ while closed ($\theta_0 = 120^\circ$) trimers maximize it.
Marginally jammed states of trimers with lower $\phi_J(\theta_0)$ exhibit quantifiably greater disorder, and the 
lower $\phi_J$ for small $\theta_0$ is apparently caused by trimers' decreasing \textit{effective} configurational freedom as they approach linearity.
\end{abstract}
\maketitle

\section{Introduction}

Identifying the densest packings of congruent particles has fascinated mankind for centuries \cite{kepler1611}.  
Computational techniques developed in recent years have facilitated identifying the densest crystalline packings of ellipsoids \cite{donev04}, the Platonic and Archimidean solids \cite{torquato09b}, ``superballs'' \cite{jiao09}, and a wide variety of convex and concave polyhedra \cite{damasceno12,torquato12}.
Such anistotropic particles have proven broadly useful since they can be assembled into structures with complex, tunable order \cite{glotzer07,torquato09}.

In contrast, dense packings of particles composed of fused spheres have received far less attention.
This is surprising because their quenched intraparticle constraints (i.e.\ the distances between and relative orientations of the fused spheres) are tunable and are not in general compatible with 3D close-packing.
For example, particles composed of 8 tangent spheres fused into a cube obviously cannot pack at $\phi_{cp} = \pi/\sqrt{18} \simeq .7405$ as individual spheres can.
Thus fused-sphere particles offer many of the same opportunities for forming solids with tunable order that those with more exotic shapes do, but with the advantage of being far easier to synthesize.

In experiments, of course, anisotropic colloidal and granular particles do not typically form bulk crystalline phases.
Special techniques are necessary to avoid jamming or glass-formation.
However, a multitude of such techniques are now available \cite{lu13,porter15}, and understanding particles' densest possible packings remains highly useful for understanding and ultimately controlling those they form under realistically achievable preparation protocols \cite{cersonsky18,torquato18}.
Thus studies that characterize both the densest packings that a given class of fused-sphere particles \textit{can} form and those that they \textit{do} form under a variety of preparation protocols, and identify key reasons for any differences between these, are of particular interest.

\begin{figure}[h]
\includegraphics[width=1.75in]{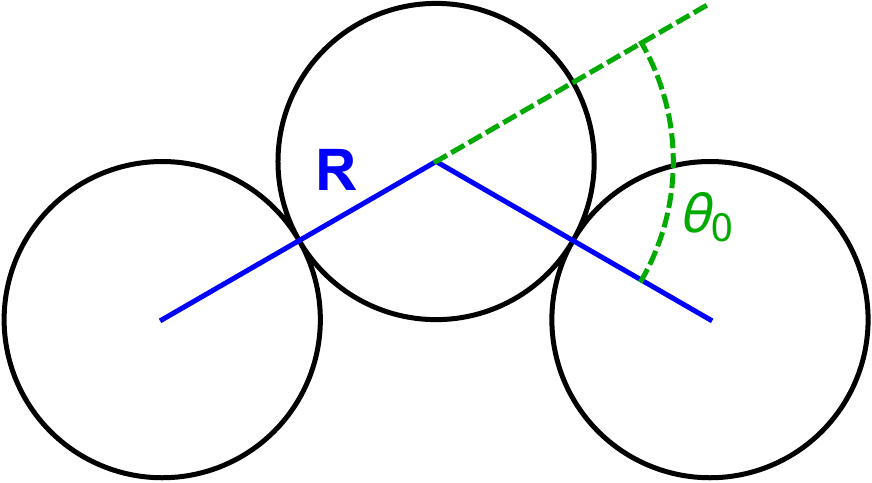}
\caption{Rigid bent-core trimers with bond angle $\theta_0$ and bond-length/monomer-diameter ratio $R$.  Here we focus primarily on the tangent-sphere ($R = 1$) case.}
\label{fig:trimermodel}
\end{figure}

Here we perform such a study for bent-core fused-sphere trimers.
As illustrated in Figure \ref{fig:trimermodel}, their shape can be characterized using two parameters: the bond angle $\theta_0$ and the ratio $R$ of intermonomer bond length to center-monomer diameter.
Tangent-sphere trimers have $R = 1$; smaller values of $R$ lead to overlap. 
While the first colloidal dimers and trimers had $R < 1$ \cite{kraft09,sacanna10,forster11}, ``colloidomers'' with $R \simeq 1$ have recently been synthesized \cite{mcmullen18,zhang18}, and  granular trimers with $R \simeq 1$ can be produced using readily available techniques such as spark welding, adhesive bonding or 3D printing \cite{olson02,scalfani14,harrington18}.
Here we will focus primarily on the tangent-sphere case because it allows for straightforward identification of the densest packings and comparison to the very extensive literature on monodisperse hard spheres.

The organization of the remainder of this paper is as follows.
In Section \ref{sec:maxd} we describe a novel method for identifying bent-core tangent-sphere trimers' densest crystalline packings, and characterize how the packings obtained via this method vary with $\theta_0$.
In Section \ref{sec:jammed} we contrast these to the disordered, jammed solid morphologies trimers form under both dynamic and quasistatic athermal compression, and also investigate the roles played by monomer overlap ($R < 1$) and spatial dimension. 
Finally, in Section \ref{sec:discuss} we summarize our results and conclude.

\section{Maximally dense packings}
\label{sec:maxd}

\subsection{Methods for obtaining crystalline structures}
\label{subsec:maxmethod}

We begin with the hypothesis that the densest crystalline packings of tangent-sphere trimers include many close-packed planes.  
This must be true for $\theta_0 = 0,\  \cos^{-1}(5/6)  \simeq 33.5573^\circ,\ 60^\circ,\ \cos^{-1}(1/3) \simeq 70.5288^\circ,\ 90^\circ,\ \rm{and}\ 120^\circ$ since trimers with these $\theta_0$ can form 3D close-packed lattices at the maximum sphere packing density $\phi_{cp}$.
It is reasonable to suppose that the densest crystals formable for other $\theta_0$ close to these six values will be slight perturbations of close-packed lattices.
Thus we employ a method that generates 3D crystals by first generating lattice planes that vary away from those found in close-packed lattices in a controlled fashion and then finding the optimal ways to stack them.

\begin{figure}[h]
\includegraphics[width=3.375in]{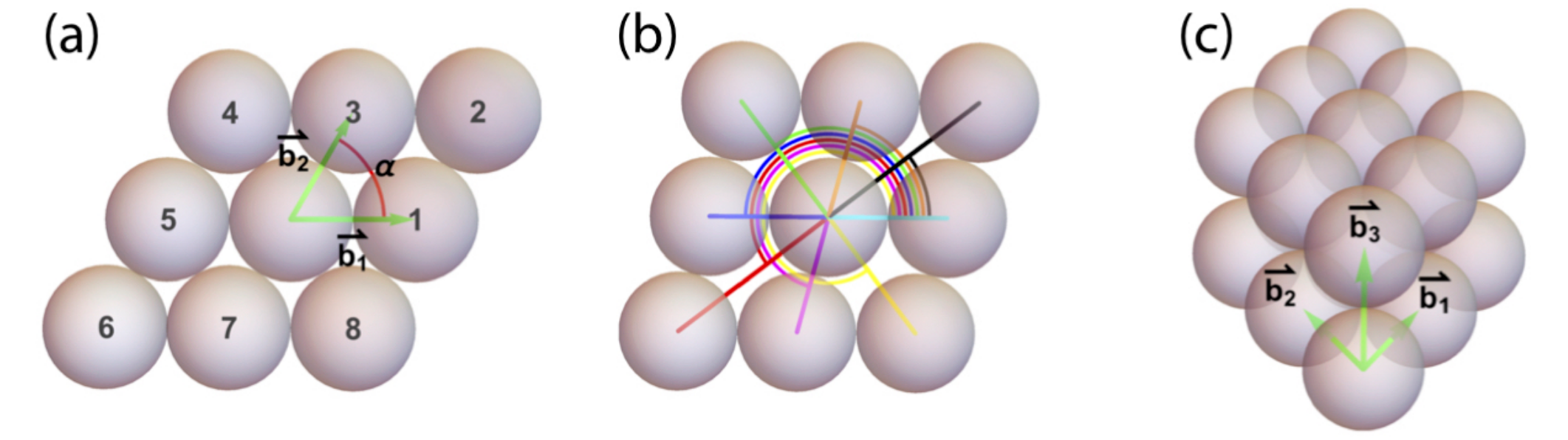}
\caption{Plane-stacking algorithm -- basic definitions. Panel (a):\ definition of $\alpha$, the angle characterizing the 2D lattice planes used to generate 3D crystals.  Panel (b):\ distances and orientation angles defining positions of spheres neighboring the reference sphere (Table \ref{tab:distor}). Panel (c):\ illustration of the vectors $\vec{b}_1,\ \vec{b}_2,\ \vec{b}_3$ that define the relative positions of nearby monomers in adjacent planes.}
\label{fig:plane}
\end{figure}

The first step is to define the planar configuration shown in Figure \ref{fig:plane}(a).
This configuration is a 2D Bravais lattice with lattice vectors $\vec{b}_1 = \{1, 0, 0\}$ and $\vec{b}_2 = \{\cos(\alpha), \sin(\alpha), 0\}$.
The positions of spheres $1-8$ are defined in Table \ref{tab:distor}.
If $\alpha = 60^\circ$, the reference sphere contacts spheres $1,\ 3,\ 4,\ 5,\ 7,\ \rm{and}\ 8$, forming the triangular lattice. 
If $\alpha = 120^\circ$, the reference sphere contacts spheres $1,\ 2,\ 3,\ 5,\ 6,\ \rm{and}\ 7$, again forming the triangular lattice.
Otherwise the reference sphere contacts spheres $1,\ 3,\ 5,\ \rm{and}\ 7$, forming a less-dense 2D lattice, e.g.\ the square lattice for $\alpha = 90^\circ$.
Now suppose [as illustrated in panel (c)] that a second identical plane is stacked above the first one, and define $\vec{b}_3$ as the vector from the center of the reference sphere in panel (a) to the center of the reference sphere in the plane above it.
We wish to solve for the $\vec{b}_3$ that will produce a maximally dense two-layer structure, i.e.\ minimize $\vec{b}_3 \cdot \hat{z}$.

\begin{figure}[h]
\includegraphics[width=2.667in]{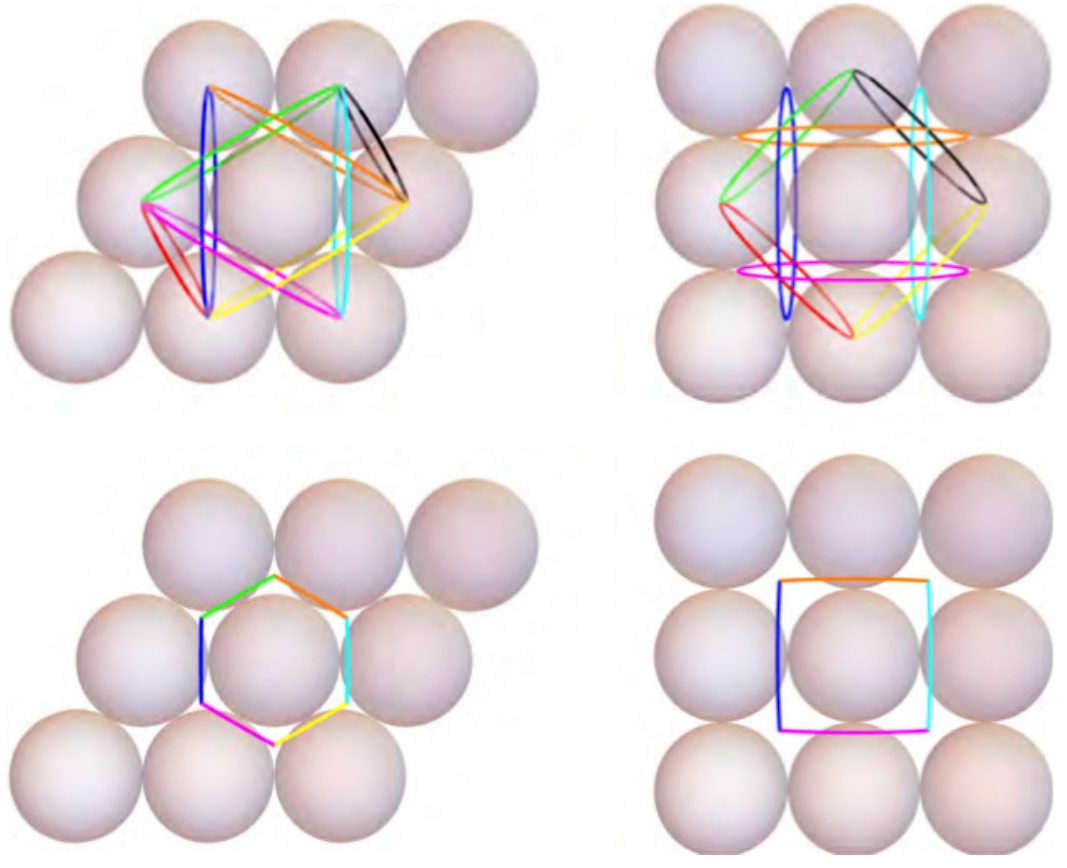}
\caption{ Plane-stacking algorithm -- circular arcs.
Top panels:\ the eight arcs associated with spheres 1-8 from Fig.\ \ref{fig:plane}(a).  
Bottom panels:\ the union of the upper portions of of these arcs (i.e.\ the portions above their $z > 0$ intersection points) is the path followed by $\vec{b}_3$ as the orientation of the second 2D layer varies.  
The left panels illustrate these features for $\alpha = 60^\circ$ while the right panels illustrate them for $\alpha = 90^\circ$.}
\label{fig:arcs}
\end{figure}

First, however, it is useful to consider arbitrarily stacked two-layer structures. 
Requiring that the layers be as close together as possible (along the $\hat{z}$ axis) for any given orientation $\mathcal{O}$ defines a unique path for $\vec{b}_3$.
To see how, consider two hard unit spheres held a fixed distance $d$ apart (with $1 \leq d \leq 2$.) 
When a third identical sphere is introduced and forced to maintain contact with both others, it will be free to rotate about the line connecting them.
The center of the third sphere will then trace out a circular disk of radius $\mathcal{R}(d) = \sqrt{1 - d^2/4}$ -- centered on and perpendicular to this line -- as it rotates.
Now suppose that the first two spheres are the reference sphere and one of the spheres $1-8$ from Fig.\ \ref{fig:plane}(a), while the third sphere lies at $\vec{b}_3$.
Hard-sphere constraints will limit the abovementioned rotation in a manner that depends only on $\alpha$;\ see Table \ref{tab:distor}, with $d = D_n$.
The accessible part of the abovementioned circular disk is a circular arc.
Figure \ref{fig:arcs} shows how combining the arcs corresponding to spheres $1-8$ yields the full set of potential configurations for the third sphere within the stacked two-layer structures we are considering.

\begin{table}[h]
\caption{Positions ($\vec{r}_n$), distances to ($D_n$), and orientations ($\mathcal{O}_n$) of spheres 1-8 in Fig.\ \ref{fig:plane} (with respect to the reference sphere) as a function of the angle $\alpha$.  The colors in the second column are illustrated in Figs.\ \ref{fig:plane}(b) and \ref{fig:arcs}.}
\begin{ruledtabular}
\begin{tabular}{ccccc}
$n$ & Color & $\vec{r}_n$ & $D_n$ & $\mathcal{O}_n$ \\ 
 1  & cyan & $\vec{b}_1$ & 1 & 0  \\ 
 2  & black & $\vec{b}_1$ + $\vec{b}_2$ & $\sqrt{2+2\cos(\alpha)}$ & $\alpha$/2  \\
 3  & orange & $\vec{b}_2$ & 1 & $\alpha$  \\
 4  & green & $\vec{b}_2 - \vec{b}_1$ & $\sqrt{2-2\cos(\alpha)}$ & ($\pi$+$\alpha$)/2  \\
 5 & blue & $-\vec{b}_1$ & 1 & $\pi$  \\
 6 & red & $-\vec{b}_1 -\vec{b}_2$ & $\sqrt{2+2\cos(\alpha)}$ & $\pi$+$\alpha$/2  \\
 7 & magenta & $-\vec{b}_2$ & 1 & $\pi$+$\alpha$  \\
 8 & yellow & $\vec{b}_1 - \vec{b}_2$ & $\sqrt{2-2\cos(\alpha)}$ & ($3\pi$+$\alpha$)/2 
\end{tabular}
\end{ruledtabular}
\label{tab:distor}
\end{table}

To describe these arcs, we will construct the vector $\vec{\ell}(d, \mathcal{O})$ extending from the origin to an arbitrary point on the edge of a circular disk that has radius $\mathcal{R}(d)$, is centered at $\vec{c}(\mathcal{O}) = d \{  \cos(\mathcal{O})/2,\ \sin(\mathcal{O})/2, 0 \}$, and is upright and faces towards the origin [i.e.\ its unit normal vector is $-\hat{c}(\mathcal{O})]$.] 
If we define a variable $\chi$ describing the angular position along the disk's rim, with $\chi = 0$ corresponding to the top of the disk, $\vec{\ell}$ can be written as
\begin{equation}
\vec{\ell}(d, \mathcal{O}, \chi) = \left[
\begin{array}{c}
d\cos(\mathcal{O})/2- \mathcal{R}(d)\sin(\mathcal{O}) \sin(\chi) \\
d\sin(\mathcal{O})/2 +\mathcal{R}(d) \cos(\mathcal{O}) \sin(\chi) \\ 
\mathcal{R}(d)\cos(\chi)
\end{array}
\right].
\label{eq:b3general}
\end{equation}
Substituting the $D_n(\alpha)$ and $\mathcal{O}_n(\alpha)$ from Table \ref{tab:distor} into this formula gives the vectors $\vec{\ell}_n(d, \mathcal{O}, \chi)$ respectively associated with spheres $n = 1-8$, i.e.\
\begin{equation}
\vec{\ell}_n(\alpha, \chi) =  \left[ \begin{array}{c}
D_n\cos(\mathcal{O}_n)/2- \mathcal{R}(D_n)\sin(\mathcal{O}_n) \sin(\chi) \\
D_n\sin(\mathcal{O})/2 +\mathcal{R}(D_n) \cos(\mathcal{O}_n) \sin(\chi) \\ 
\mathcal{R}(D_n)\cos(\chi)
\end{array} \right].
\label{eq:b3simple}
\end{equation}
Since $D_n$ and $\mathcal{O}_n$ are functions only of $\alpha$, the vectors $\vec{\ell}_n$ are functions of only two variables:\ $\alpha$ and $\chi$.
The various circles shown in the top panels of Fig.\ \ref{fig:arcs} are traced out by $\vec{\ell}_n(\alpha,\chi)$ as $\chi$ varies from $0$ to $2\pi$.

\begin{figure}[h]
\includegraphics[width=3.125in]{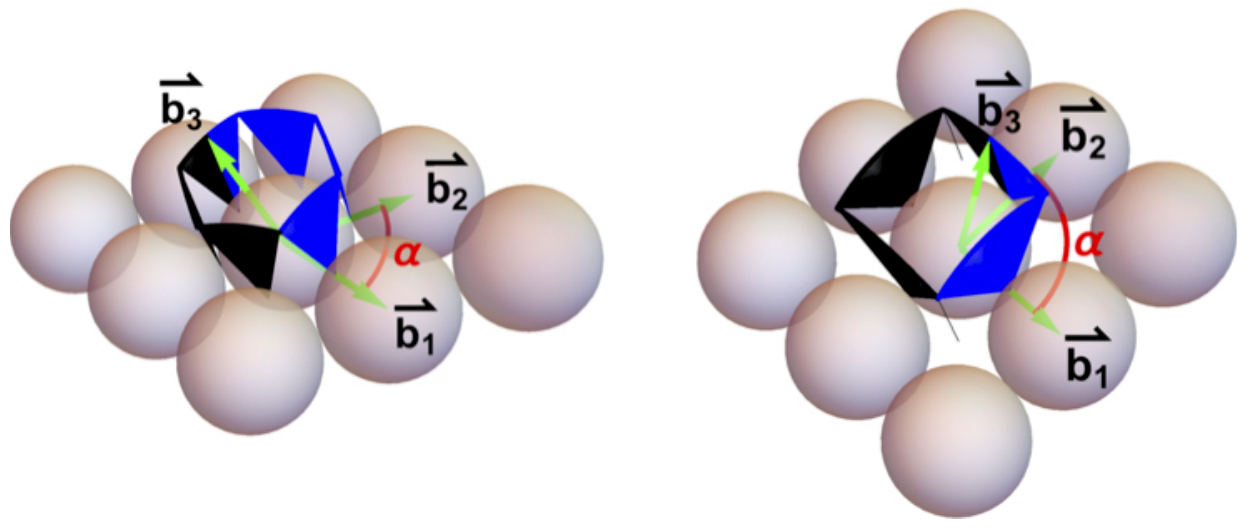}
\caption{Plane-stacking algorithm -- depiction of the traversal made by $\vec{b}_3$ as $\beta$ varies.  The configuration of the reference layer and subtended circular arcs are shown for [left panel:\ $\alpha = 60^\circ,\ \beta = 7/12$] and [right panel:\ $\alpha = 90^\circ,\ \beta = 3/8$].  The blue-shaded regions have been traversed for these values of $\beta$ while the gray-shaded regions have yet to be traversed.} 
\label{fig:b3betaschem}
\end{figure}

The next step is to identify the ranges of $\chi$ corresponding to the abovementioned circular arcs.
Finding the arcs' points of intersection (as a function of $\alpha$) will allow us to determine the path taken by $\vec{b}_3$ as the upper layer in Fig.\ \ref{fig:plane}(c) is shifted around at the minimum height that respects the hard-sphere nonoverlap constraints.
It turns out that these points can be described using only two additional scalar functions $\chi_1(\alpha)$ and $\chi_2(\alpha)$, neither of which depend on $n$:
\begin{equation}
\begin{array}{c}
\chi_1(\alpha) = \displaystyle\frac{\pi}{2} - \sin^{-1}\left[ \sqrt{\displaystyle\frac{2[1 + 2\sin(q)]}{3[1 + \sin(q)]}} \right] \\
\\
\chi_2(\alpha) =  \displaystyle\frac{\pi}{2}-2\sin^{-1}\left(\displaystyle\frac{1}{\sqrt{2+2\sin{q}}}\right)
\end{array},
\label{eq:chi12}
\end{equation}
where $q \equiv |\alpha - \pi/2|$.
The total range of $\chi$ subtended by the circular arcs as the upper layer traverses the gaps is
\begin{equation}
\Delta(\alpha) = 8\chi_1(\alpha)+4\chi_2(\alpha).
\label{eq:deltaofalpha}
\end{equation}
If we define a parameter $\beta$ that varies from $0$ to $1$ during a full traversal, the vector $\vec{b}_3(\alpha,\beta)$ is given by the formulas listed in Table \ref{tab:b3pieces}.
As is apparent from Fig.\ \ref{fig:arcs}, the vector $\vec{\ell}$ passes over at most six of the arcs corresponding to spheres $1-8$ as it traverses the gaps; the others are excluded because their $D_n$ are too large.
These six arcs are labeled $i = 1,\ 2,\ ...,\ 6$ in the table, and the traversal of $\vec{b}_3$ over them
is schematically depicted in Figure \ref{fig:b3betaschem}.

\begin{table}[h!]
\caption{Definition of $\vec{b}_3$ in terms of the quantities defined in Eqs.\ \ref{eq:b3simple}-\ref{eq:deltaofalpha}:\ $\vec{b}_3 = \vec{b}_3^i(\alpha, \beta)$ when $\beta_{min}^i \leq \beta \leq \beta_{max}^i$.}
\begin{ruledtabular}
\scriptsize
\begin{tabular}{cccc}
& & $60^\circ \leq \alpha \leq 90^\circ$ & \\
\hline
$i$ & $\Delta(\alpha)\beta_{min}^i$ & $\Delta(\alpha)\beta_{max}^i$ & $\vec{b}_3^i(\alpha, \beta)$\\ 

1 & 0 & $2\chi_1(\alpha)$ & $\vec{\ell}_1[\alpha, \beta\Delta(\alpha) - \chi_1(\alpha)]$ \\

2 & $2\chi_1(\alpha)$ & $4\chi_1(\alpha)$ & $\vec{\ell}_3[\alpha, \beta\Delta(\alpha) - 3\chi_1(\alpha)]$ \\

3 & $4\chi_1(\alpha)$ & $4\chi_1(\alpha) + 2\chi_2(\alpha)$ & $\vec{\ell}_4[\alpha, \beta\Delta(\alpha) - 4\chi_1(\alpha) - \chi_2(\alpha)]$ \\

4 & $4\chi_1(\alpha) + 2\chi_2(\alpha)$ & $6\chi_1(\alpha) + 2\chi_2(\alpha)$ & $\vec{\ell}_5[\alpha, \beta\Delta(\alpha) - 5\chi_1(\alpha) - 2\chi_2(\alpha)]$\\

5 & $6\chi_1(\alpha) + 2\chi_2(\alpha)$ & $8\chi_1(\alpha) + 2\chi_2(\alpha)$ & $\vec{\ell}_7[\alpha, \beta\Delta(\alpha) - 7\chi_1(\alpha) - 2\chi_2(\alpha)]$\\

6 & $8\chi_1(\alpha) + 2\chi_2(\alpha)$ &  $8\chi_1(\alpha) + 4\chi_2(\alpha)$ & $\vec{\ell}_8[\alpha, \beta\Delta(\alpha) - 8\chi_1(\alpha) - 3\chi_2(\alpha)]$ \\

\hline
& & $90^\circ \leq \alpha \leq 120^\circ$ & \\
 \hline
$i$ & $\Delta(\alpha)\beta_{min}^i$ & $\Delta(\alpha)\beta_{max}^i$ & $\vec{b}_3^i(\alpha, \beta)$\\

1 & 0 & $2\chi_1(\alpha)$ & $\vec{\ell}_1[\alpha, \beta\Delta(\alpha) - \chi_1(\alpha)]$ \\

2 & $2\chi_1(\alpha)$ & $2\chi_1(\alpha) + 2\chi_2(\alpha)$ & $\vec{\ell}_2[\alpha, \beta\Delta(\alpha) - 2\chi_1(\alpha) - \chi_2(\alpha)]$ \\

3 & $2\chi_1(\alpha) + 2\chi_2(\alpha)$ & $4\chi_1(\alpha) + 2\chi_2(\alpha)$ & $\vec{\ell}_3[\alpha, \beta\Delta(\alpha) - 3\chi_1(\alpha) - 2\chi_2(\alpha)]$ \\

4 & $4\chi_1(\alpha) + 2\chi_2(\alpha)$ & $6\chi_1(\alpha) + 2\chi_2(\alpha)$ & $\vec{\ell}_5[\alpha, \beta\Delta(\alpha) - 5\chi_1(\alpha) - 2\chi_2(\alpha)]$ \\

5 & $6\chi_1(\alpha) + 2\chi_2(\alpha)$ & $6\chi_1(\alpha) + 4\chi_2(\alpha)$ & $\vec{\ell}_6[\alpha, \beta\Delta(\alpha) - 6\chi_1(\alpha) - 3\chi_2(\alpha)]$ \\

6 & $6\chi_1(\alpha) + 4\chi_2(\alpha)$ & $8\chi_1(\alpha) + 4\chi_2(\alpha)$ &  $\vec{\ell}_7[\alpha, \beta\Delta(\alpha) - 7\chi_1(\alpha) - 4\chi_2(\alpha)]$ 
\end{tabular}
\normalsize
\end{ruledtabular}
\label{tab:b3pieces}
\end{table}

3D space can be filled with unit-diameter spheres at positions $\vec{r}_{uvw}(\alpha, \beta) = u\vec{b}_1 + v\vec{b}_2(\alpha) + w\vec{b}_3(\alpha, \beta)$, where $\{ u, v,  w \} \in \mathbb{Z}^3$, but the resultant lattices will be only a small subset of those we need to consider to identify the maximally-dense trimer crystals for all $\theta_0$.
However, arbitrary space-filling planar stackings can be described using $M$-plane bases where the positions of successive planes are related using $M$ different $\beta$ values $\{ \beta_1, ..., \beta_M \}$.
We constrain these planes to have the same value of $\alpha$ since different $\alpha$ would produce massive incommensurability.
Table \ref{tab:M2} illustrates the diversity achievable with this method for $M = 2$.
It also illustrates a key strength of our method:\ the ability to generate different lattice planes of the same crystal by varying $\alpha$.
For example, the planes for $\alpha = 60^\circ\ \textrm{or}\ 120^\circ$ correspond to the $\{ 1\ 1\ 1 \}$ planes of an FCC lattice, whereas for $\alpha = 90^\circ$ they correspond to the $\{ 1\ 0\ 0 \}$ planes of the same FCC lattice.

\begin{table}[h!]
\caption{Example crystalline orderings for $M = 2$.}
\begin{ruledtabular}
\footnotesize
\begin{tabular}{ccc}
$\alpha (^\circ)$ & Condition for $\beta_1, \beta_2$ & Order \\
$60$ or $120$ & $(\beta_1\vee\beta_2\bmod\frac{1}{6}=0) \wedge (|\beta_1-\beta_2|\bmod\frac{1}{3}=0)$  & FCC \\
$60$ or $120$ & $(\beta_1\vee\beta_2\bmod\frac{1}{6}=0) \wedge  (|\beta_1-\beta_2|\bmod\frac{1}{3}=\frac{1}{6})$ & HCP\\
$90$ & $(\beta_1\vee\beta_2\bmod\frac{1}{4}=0) \wedge (|\beta_1-\beta_2|\bmod\frac{1}{4}=0)$ & FCC
\end{tabular}
\normalsize
\end{ruledtabular}
\label{tab:M2}
\end{table}

Trimers with any $\theta_0$ can form periodic ($M \geq 2$)-layer planar stackings where each periodic group includes at least a bilayer of close-packed crystal [as in Fig.\ \ref{fig:plane}(c)].
If $\theta_0$ is incommensurable with 3D close-packing, trimer crystals cannot include more than three consecutive close-packed layers; they must include ``defects.''
Results from 2D systems \cite{griffith18} suggest that these will take the form of ``gaps'' between close-packed bilayers or trilayers that close when $\theta_0$ \textit{is} commensurable with 3D close-packing, and hence that maximally-dense bent-core trimer crystals for all $\theta_0$ can be identified using only $2 \leq M \leq 3$.

Rather than identifying the densest trimer crystals using a linear-programming or Monte-Carlo-like algorithm such as Torquato \& Jiao's ASC \cite{torquato10,atkinson12}, we identify them using our $\alpha-\beta$ formalism.
The vectors $\vec{b}_1,\ \vec{b}_2(\alpha),\ \rm{and}\ \vec{b}_3(\alpha,\beta)$ define a family of parallelopipeds.
Consider a fundamental cell $\mathcal{C}$ containing $n_x \times n_y \times n_z$ such parallelopipeds along the x, y, and z directions.
The cell can be fully described as
\begin{equation}
\mathcal{C} = \mathcal{C}(n_x, n_y, n_z, \alpha, \beta_1, ..., \beta_M).
\label{eq:Cgenl}
\end{equation}
$\mathcal{C}$ tiles space and forms the basis for a periodic $M$-layer planar stacking if $n_z = M$.
Since $\mathcal{C}$ must contain an integer number of trimers ($n_{tri}$), we have the restriction $n_x n_y M = 3n_{tri}$.
Results for 2D bent-core tangent-disk trimers \cite{griffith18} as well as other concave hard particles with a wide variety of shapes \cite{torquato12} suggest that the densest packings for arbitrary $\theta_0$ will be Kuperberg double lattices \cite{kuperberg90} and hence that $n_{tri} = 2$ is a sufficiently large basis.

If these crystals consist of close-packed $M$-layers separated by gap layers, we can set all but one of the $\{ \beta \}$ to zero, i.e.\ we can set $\beta_1 = 0$ for $M = 2$ and $\beta_1 = \beta_3 = 0$ for $M = 3$.
The fundamental cells can now be fully described as
\begin{equation}
\mathcal{C} = \mathcal{C}(n_x, n_y, M, \alpha, \beta_2).
\label{eq:Cgenl2}
\end{equation}
We find putatively maximally-dense trimer crystals by looping over the remaining variables $\{ n_x, n_y, \alpha, \beta_2 \}$.
For $n_{tri} = 2$ the loops over $n_x$ and $n_y$ are trivial: $\{ n_x, n_y \} = \{ (1,2),  (2,1) \}$ for $M =3$ and  $\{ n_x, n_y \} = \{ (1,3),  (3,1) \}$ for $M =2$.
Since all intermonomer distances are invariant under the transformation $\vec{b}_3(90^\circ - p, \beta) \to \vec{b}_3(90^\circ + p, \beta)$ for all $0 \leq p \leq 30^\circ$, we consider the ranges $60^\circ \leq \alpha \leq 90^\circ$ and $0 \leq \beta_2 \leq 1$.
The loop we execute is a double loop over all $(\alpha_s, \beta_{2,t})$ where $\alpha_s = (60 + .12s)^\circ$ and $\beta_{2,t} = t/600$, where $s \in [0,250]$ and $t \in [200,400]$.
Each $(s, t)$ pair produces a different periodic planar stacking, so the total number of distinct packings produced is $50451$ \cite{famIVnote0}.

Careful readers will note that we have not yet determined which of these packings correspond to \textit{trimer} crystals.
We do this by identifying all contacting $(i, j, k)$ and $(l, m, n)$ triplets within the fundamental cells that consist of spheres at positions $(\vec{r}_i, \vec{r}_j, \vec{r}_k, \vec{r}_l, \vec{r}_m, \vec{r}_n)$ that satisfy $r_{ij} = r_{jk} = 1$ and  $r_{lm} = r_{mn} = 1$, where $\vec{r}_{ij} = \vec{r}_j - \vec{r}_i$ and so on, and of course $i \neq j \neq k \neq l \neq m \neq n$.
During this procedure, periodic boundary conditions are applied to the fundamental cell along all three directions and the minimum image convention is employed.
The $(i, j, k)$ and $(l, m, n)$ triplets identified this way have bond angles $\theta_{ijk} = \cos^{-1}(\vec{r}_{ij}\cdot\vec{r}_{jk})$ and $\theta_{lmn} = \cos^{-1}(\vec{r}_{lm}\cdot\vec{r}_{mn})$, respectively.
If $\theta_{ijk} = \theta_{lmn}$, then the packing is a bent-core trimer crystal with $\theta_0 = \theta_{ijk} = \theta_{lmn}$.
We store its packing fraction in the set $\{ \phi_{st}(\theta_0) \}$, where again $s \in [0,250]$ and $t \in [200,400]$.
This procedure allows us to identify $\phi_{max}(\theta_0) = \max( \{ \phi_{st}(\theta_0) \} )$ for \textit{all} $\theta_0$ in a single sweep.
Note that its essential feature is that we impose the fixed-angle $(\theta = \theta_0)$ restriction \textit{a posteriori} rather than \textit{a prori}. 
We find that the above procedure is far more efficient for these systems than alternative methods such as ASC.

\subsection{Densest tangent-sphere-trimer crystals}
\label{subsec:dtc}

\begin{figure}[h]
\includegraphics[width=3.25in]{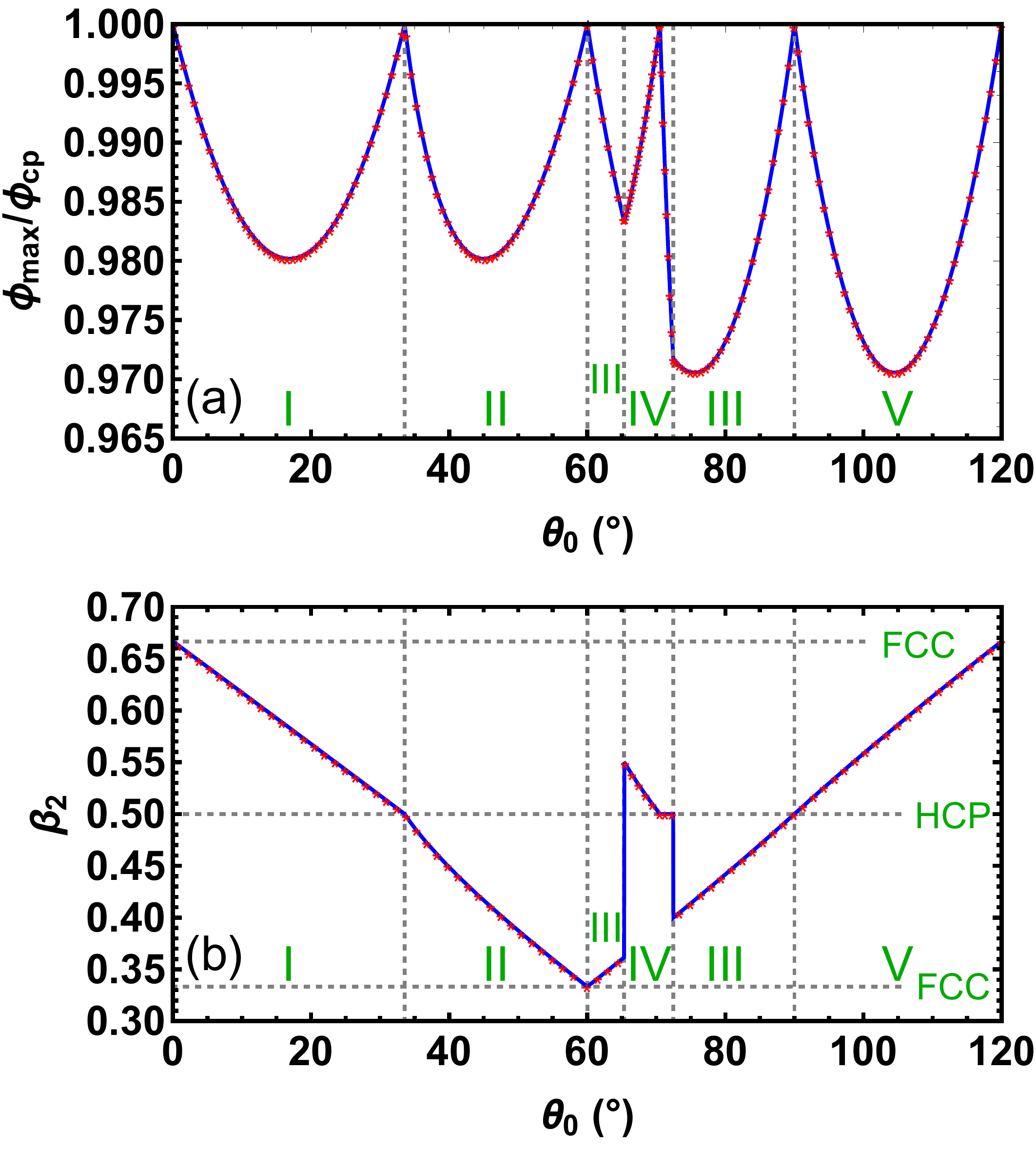}
\caption{Structural order of maximally dense bent-core tangent-sphere trimer crystals.  In both panels, the blue curves show analytic results from Table \ref{tab:phimaxanalyt} while red symbols show numerical results from the method described in Section \ref{subsec:maxmethod}.  The ``FCC/HCP'' horizontal lines in panel (b) describe the $\beta_3(\theta_0)$-dependent order when $\alpha(\theta_0) = 60^\circ$ (as is the case for all families except IVb.)}
\label{fig:phimax}
\end{figure}

Numerical results for $\phi_{max}(\theta_0)$ are shown with red symbols in Figure \ref{fig:phimax}(a).
$\phi_{max}(\theta_0)$ is apparently a piecewise-smooth function with either five or six distinct branches.
Four of these branches begin and end with close-packed crystals.
The smooth variation of $\phi_{max}$ between cusplike maxima at the $\theta_0$ that are commensurable with close-packing is reminiscent of other packing problems where $\phi_{max}(p)$ varies smoothly between cusplike maxima as a function of some grain-shape or confinement-geometry parameter $p$ \cite{torquato18,mughal11,mughal12}, and strongly suggests that the crystals lying on each branch share many common features.
We therefore label these branches as ``families'' I, II, III, and V.
While the other branch (family IV) does not \textit{obviously} begin/end with a close-packed crystal, it does include one [at $\theta_0 = \cos^{-1}(1/3)$], suggesting that its crystals also share a common character.

\begin{table}[h]
\caption{Ordering of maximally-dense bent-core tangent-sphere trimer crystals -- basic properties.   Here $\theta_1 = 65.3512^\circ$ and $\theta_2 = 72.4530^\circ$.  The average monomer coordination number $Z$ includes both covalent bonds and noncovalent contacts.  Note that $Z(\theta_0) = 12$ for $\theta_0 = 0,\  \cos^{-1}(5/6),\ 60^\circ,\ \cos^{-1}(1/3),\ 90^\circ,\ \rm{and}\ 120^\circ$; the values given below are for the intermediate $\theta_0$.}
\begin{ruledtabular}
\begin{tabular}{ccccc}
Family & Range of $\theta_0$ & $n_{tri}$ & $M$ & $Z$ \\
I & 0 - $\cos^{-1}(5/6)$  & 1 & 3 & $11 \frac{1}{3}$ \\
II & $\cos^{-1}(5/6)$ - $60^\circ$ & 1 & 3 & $11 \frac{1}{3}$ \\
III & $60^\circ$ - $\theta_1$ , $\theta_2$ - $90^\circ$ & 2 & 2 & 11 \\
IVa & $\theta_1$ - $\cos^{-1}(1/3)$ & 1 & 3 &  $11 \frac{1}{3}$ \\
IVb &  $\cos^{-1}(1/3)$ -  $\theta_2$ & 1 & 3 & $10$ \\
V & $90^\circ$ - $120^\circ$ & 2 & 2 & $11$ \\
\end{tabular}
\end{ruledtabular}
\label{tab:genprops}
\end{table}

\begin{figure}[h]
\includegraphics[width=2.6in]{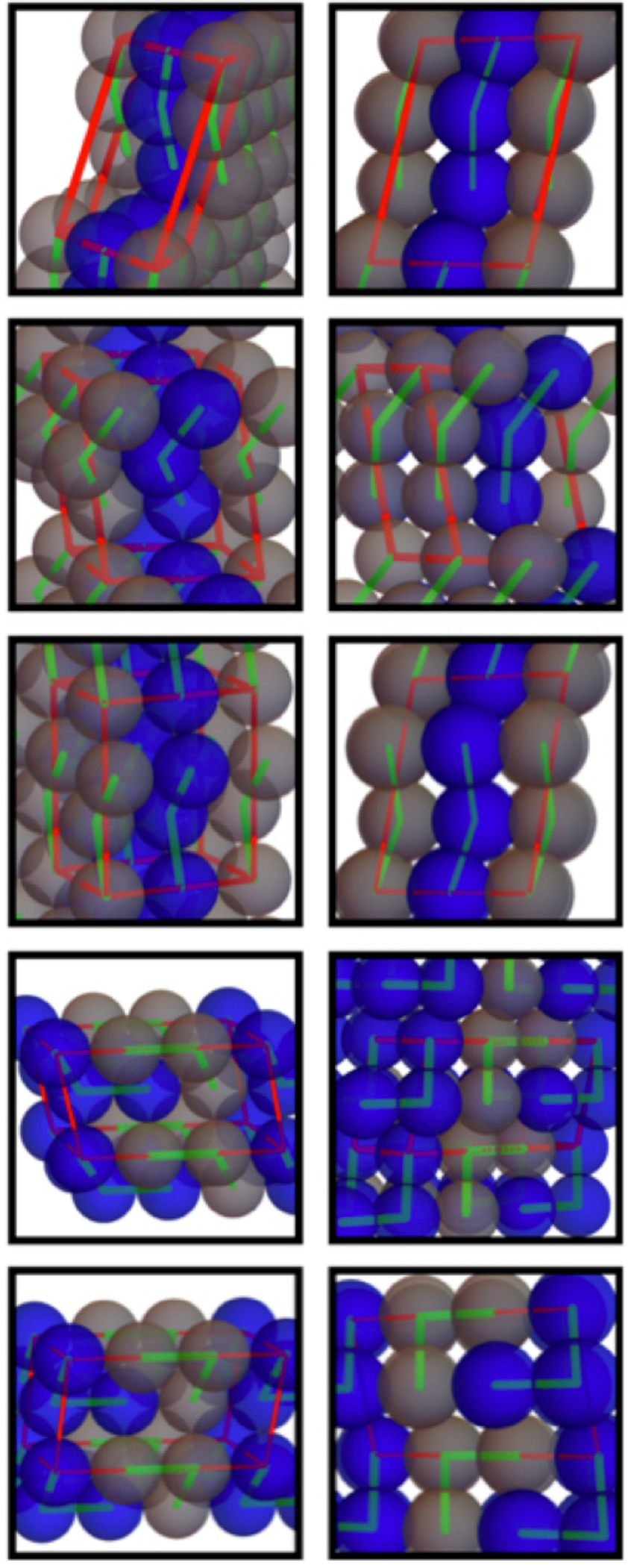}
\caption{Illustration of bent-core trimer crystals' $\theta_0$-dependent structure.  Panels in the left column highlight the fundamental cells $\mathcal{C}$ for the least dense members of families I-III, IVa, and V, i.e.\ the cells for $\theta_0 = 16.84^\circ,\ 45^\circ,\ 65.36^\circ,\ 75.52^\circ,\ \rm{and}\ 104.47^\circ$ (from top to bottom).  Panels in the right column highlight the close-packed-bi/trilayer-plus-gap structure for the same systems.  In all panels, red lines indicate the boundaries of $\mathcal{C}$ and green lines indicate covalent backbone bonds.}
\label{fig:unitcells}
\end{figure}

Each family's characteristic unit-cell and gap-layer structure is depicted in Figure \ref{fig:unitcells}, and Table \ref{tab:genprops} further summarizes their basic properties.
All the crystals within any given family share common values of of $n_{tri}$ and $M$.
As hypothesized above, for $\theta_0 \leq 60^\circ$ the densest bent-core-trimer crystals have a single-trimer basis and consist of close-packed trilayers separated by gap layers.
For $\theta_0 > \theta_2$ they are Kuperberg double lattices \cite{kuperberg90} consisting of two interpenetrating lattices of trimers related by a displacement plus a $180^\circ$ rotation of all constituents about their centers of inversion symmetry, and have close-packed \textit{bilayers} separated by gap layers.
In both cases, the size of the gaps increases with the distance of $\theta_0$ from angles commensurable with close-packing.
The reason for the trilayer-vs-bilayer distinction is the same as for 2D bent-core trimers \cite{griffith18} -- the inability of a reference trimer to form a bond-triangle on its concave side with a monomer belonging to a second trimer when $\theta_0 > 60^\circ$.
This distinction quantitatively predicts the degree to which the minima of $\phi_{max}(\theta_0)$ are lower for families III and V than for families I and II.
The minimal densities of families I, II, III and V are respectively $\phi_1^* = .980181\phi_{cp}$ (at $\theta_0 = 16.8421^\circ$), $\phi_2^* = \phi_1^*$ (at $\theta_0 = 45^\circ)$, $\phi_3^* = .970563\phi_{cp}$ (at $\theta_0 = 75.5225^\circ$), and $\phi_5^* = \phi_3*$ (at $\theta_0 = 104.478^\circ$).
They satisfy 
\scriptsize
\begin{equation}
\displaystyle\frac{\phi_{cp}-\phi_3^*}{\phi_{cp}-\phi_1^*} \cdot \displaystyle\frac{\phi_1^*}{\phi_3^*} = \displaystyle\frac{\phi_{cp}-\phi_5^*}{\phi_{cp}-\phi_1^*} \cdot \displaystyle\frac{\phi_1^*}{\phi_5^*} = \displaystyle\frac{\phi_{cp}-\phi_3^*}{\phi_{cp}-\phi_2^*} \cdot \displaystyle\frac{\phi_1^*}{\phi_2^*} = \displaystyle\frac{\phi_{cp}-\phi_5^*}{\phi_{cp}-\phi_2^*} \cdot \displaystyle\frac{\phi_2^*}{\phi_5^*} = \displaystyle\frac{3}{2}.
\end{equation}
\normalsize
Here the $3/2$ reflects the fact that the gap-layer density is $50\%$ higher for families III and V than it is for families I and II.
Similar relations hold for 2D systems \cite{griffith18}.

\begin{table*}
\caption{Ordering of the densest bent-core tangent-sphere trimer crystals -- analytic expressions for $\beta_2(\theta_0)$ and $\phi_{max}(\theta_0)$.  All families have $\alpha(\theta_0) = 60^\circ$ except for family IVb, which has $\alpha(\theta_0) = \sec^{-1}\left[ \sec(\theta_0) - 1 \right]$.}
\begin{ruledtabular}
\scriptsize
\begin{tabular}{ccc}
Family &  $\beta_2(\theta_0)$ & $\phi_{max}(\theta_0)/\phi_{cp}$ \\

I & $\displaystyle\frac{2}{3} + \displaystyle\frac{ \cos^{-1}\left(\frac{1}{3}\left[ 4\cos(\theta_0) - 1\right] \right) }{6\left[ 180^\circ - 4
\sec^{-1}(\sqrt{3}) \right]}$ & $\displaystyle\frac{9}{5 + 4\cos(\theta_0) + \sqrt{\cos(\theta_0) - \cos(2\theta_0)} }$ \\

II & $\displaystyle\frac{5}{12} + \displaystyle\frac{\sin^{-1}\left[ \frac{1}{3} \left( 2\sqrt{2}\cos(\theta_0) + \sqrt{1 - 2\cos(2\theta_0)} \right) \right] \rm{sign}(\pi/4 - \theta_0) }{ 6\left[ \pi - 4\sec^{-1}(\sqrt{3}) \right] }$  & $\displaystyle\frac{12}{ 4[2+\cos(\theta_0)] + \sqrt{2 - 4\cos(2\theta_0) } }$ \\

III & $\displaystyle\frac{5}{12} - \displaystyle\frac{  \sin^{-1}\left(\frac{1}{3}\left[ 4\cos(\theta_0) - 1\right] \right)  }{ 6\left[ 180^\circ - 4\sec^{-1}(\sqrt{3}) \right] }$ & $\displaystyle\frac{2}{1 +  \sqrt{\cos(\theta_0) - \cos(2\theta_0)} }$ \\

IVa & $\displaystyle\frac{5}{12} + \displaystyle\frac{ \sin^{-1} \left( \frac{1}{9} \left[ 4\cos(\theta_0) - 8\sqrt{\cos(\pi - \theta_0) + \cos(\pi - 2\theta_0) } + 1 \right] \right) }{ 6\left[ 180^\circ - 4\csc^{-1}(\sqrt{3}) \right]  } $ & $\displaystyle\frac{18\sqrt{2}}{ 12\sqrt{2} + \sqrt{ 81 - \left( 1 + 4\cos(\theta_0) - 8\sqrt{ \cos(180^\circ - \theta_0) - \cos(180^\circ - 2\theta_0) } \right)^2 } }$ \\

IVb & $1/2$  & $\displaystyle\frac{\sqrt{2}\sin(\theta_0/2)\tan(\theta_0/2)}{\sqrt{1 - 2\cos(\theta_0)}}$ \\

V &  $\displaystyle\frac{11}{12} + \displaystyle\frac{ \sin^{-1}\left( \frac{1}{3} \left[ 4\cos(\theta_0) + 1 \right]  \right)}{ 6\left[ 180^\circ - 4\sec^{-1}(\sqrt{3}) \right] }$ & $\displaystyle\frac{2}{ 1 +  \sqrt{\cos(180^\circ - \theta_0) + \cos(180^\circ - 2\theta_0)} }$
\end{tabular}
\normalsize
\end{ruledtabular}
\label{tab:phimaxanalyt}
\end{table*}

Additional insight into the structure of these maximally-dense crystals can be gained by examining the topology of their bond/contact network and trends in their $\beta_2(\theta_0)$.
Both the close-packed-trilayer-plus-gap families (I, II, and IVa) and the close-packed-bilayer-plus-gap families (III and V) share the same average monomer coordination number $Z$ in addition to the same $\phi^*$.
Numerical results for $\beta_2(\theta_0)$ are given in Fig.\ \ref{fig:phimax}(b).
These results, together with visual inspection of the fundamental cells and gap-layer structure, suggests that the evolution of these families' structure as $\theta_0$ varies can be described (mathematically) as a continuous displacive transformation.
All our results are consistent with the hypothesis that these are Shoji-Nishiyama-like FCC$\leftrightarrow$HCP shear transformations \cite{cayron16}.
Family IVb is an exception to this pattern, for reasons we will discuss below.

As discussed above, all these crystals are characterized by their fundamental cells $\mathcal{C}(\theta_0) = \mathcal{C}(n_x, n_y, n_z, \alpha, \beta_2)$ (Eqs.\ \ref{eq:Cgenl}-\ref{eq:Cgenl2}), where each of the arguments to the $\mathcal{C}$ function are $\theta_0$-dependent.
Recall that $\mathcal{C}$ is a parallelopiped defined by the vectors $n_x \vec{b}_1$, $n_y \vec{b}_2(\alpha)$, and $(M-1)\vec{b}_3(\alpha,0) + \vec{b}_3(\alpha,\beta_2)$.
Its associated packing fraction is simply
\begin{equation}
\begin{array}{lcl}
\phi[\mathcal{C}] & = & \displaystyle\frac{ 3n_{tri} (\pi/6) }{ [ n_x  \vec{b}_1 \times n_y  \vec{b}_2(\alpha) ]\cdot [ (M-1)\vec{b}_3(\alpha,0) + \vec{b}_3(\alpha,\beta_2) ] } \\
\\
& = &  \displaystyle\frac{ \pi n_{tri} }{ 2 n_x n_y \sin(\alpha) \hat{z} \cdot  [ (M-1)\vec{b}_3(\alpha,0) + \vec{b}_3(\alpha,\beta_2) ]  }
\end{array}.
\label{eq:phiC}
\end{equation}
The numerical data shown in Fig.\ \ref{fig:phimax}(b) suggest that each family's $\beta_2(\theta_0)$ is describable by a continuous analytic function.
This is indeed the case; expressions for the $\beta_2(\theta_0)$ are given in Table \ref{tab:phimaxanalyt}.
Since the densest crystals' $\alpha(\theta_0)$ are also given by exact analytic expressions, we now have a complete set of parameters and analytic functions to plug into Equation \ref{eq:phiC}.
Plugging in values of $n_{tri}$ and $M$ (Tab.\ \ref{tab:genprops}) together with these analytic forms for $\alpha(\theta_0)$ and $\beta_2(\theta_0)$ yields the exact expressions for $\phi_{max}(\theta_0)$ given in Tab.\ \ref{tab:phimaxanalyt}.
Note that the values of $\theta_1$ and $\theta_2$ given in Tab.\ \ref{tab:genprops} were obtained by solving for the $\theta_0$ at which the relevant analytic $\phi_{max}(\theta_0)$ functions are equal.

\begin{figure}[h]
\includegraphics[width=3.25in]{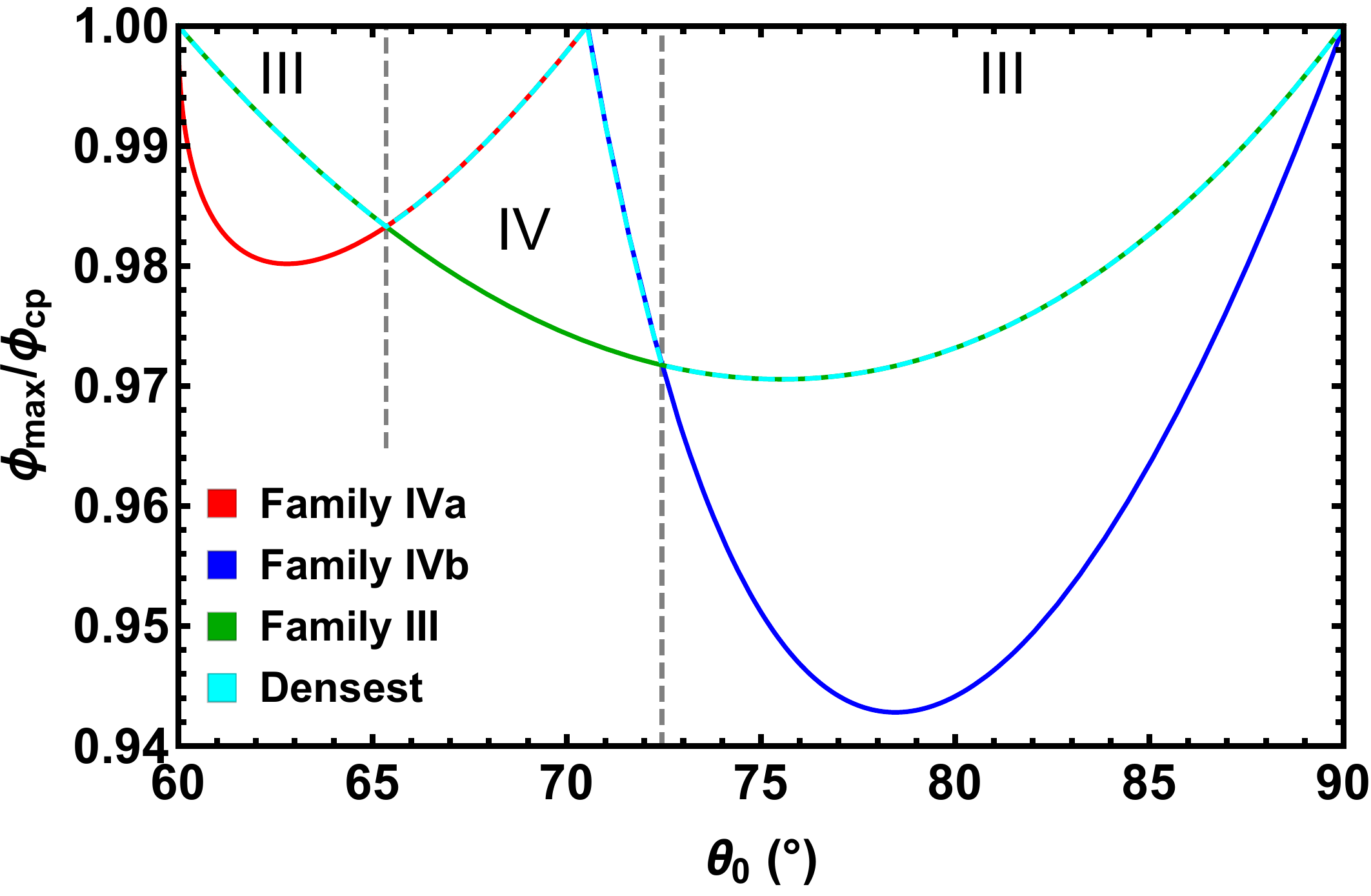}
\caption{Maximally and nonmaximally dense crystalline structures in the range $60^\circ \leq \theta_0 \leq 90^\circ$.  The curves show the analytic functions given in Table \ref{tab:phimaxanalyt}.}
\label{fig:60to90}
\end{figure}

Careful readers will note that we have not yet discussed family IV's structure in detail; we do so now.
Clearly the solution presented in Fig.\ \ref{fig:phimax} and Tab.\ \ref{tab:genprops} is more complicated in the range $60^{\circ} \leq \theta_0 \leq 90^\circ$ than it is for $\theta_0 < 60^\circ$ or $\theta_0 > 90^\circ$.
The dashed cyan curve in Figure \ref{fig:60to90} shows $\phi_{max}(\theta_0)$ for families III and IV.
Family III is the densest structure for two disjoint ranges of $\theta_0$: $60^\circ \leq \theta_0 \leq \theta_1$ and $\theta_2 \leq \theta_0 \leq 90^\circ$.
The reason for the abovementioned disjointness is simply that family IV's density is greater than family III's in the range $\theta_1 \leq \theta_0 \leq \theta_2$.
Family IVa's close-packed-trilayer-plus-gap structure is similar to families I and II; its minimal density $\phi_{4a}^* = \phi_1^* = \phi_2^*$ occurs at $\theta_0 = 62.80^\circ$.
Family IVb also has a trilayer-plus-gap structure, but is unique among families I-V in that [except at $\theta_0 = \cos^{-1}(1/3)$] its $\alpha(\theta_0) \neq 60^\circ$ and hence its trilayers are not close-packed. 
It is also unique in having a constant $\beta_2(\theta_0) = 1/2$; note that the distance between the stacked planes depicted in Figs.\ \ref{fig:plane}-\ref{fig:b3betaschem} is minimal for $\beta = 0\ \rm{and}\ 1/2$.
Thus, in contrast to families (I-IVa, V) where gap layers accommodate the incommensurability of $\theta_0$ with 3D close-packing, family IVb maximizes density while accomodating this incommensurability by adopting a stacked-plane structure that is not triangular.

All of the above discussion has omitted one key consideration:\ degeneracy of the solutions for $\beta_2$.
For example, the fact that $\theta_0 = 0,\ 60^\circ,\ 90^\circ,\ \rm{and}\ 120^\circ$ trimers can form both FCC and HCP crystals while $\theta_0 = \cos^{-1}(5/6)\ \rm{and}\ \cos^{-1}(1/3)$ trimers can form HCP but not FCC crystals is not apparent from Fig.\ \ref{fig:phimax}(b).
When $\alpha = 60^\circ$, the structure of trimer crystals is invariant under shifts of $\beta_2(\theta_0)$ by $\pm 1/3$ and also under reflections of $\beta_2(\theta_0)$ about $\beta_2 = 0,\ 1/6,\ 1/3,\ 1/2,\ 2/3,\ 5/6,\ \rm{and}\ 1$.
A second type of degeneracy arises for $\theta_0 \geq 60^\circ$:\ reflections of  $\beta_2(\theta_0)$ about $m/12$, where $m$ is an odd integer.
These operations leave $\phi$ unchanged but \textit{do} change crystals' structure.
For example, the reflection  $\beta_2(\theta_0) \rightarrow \beta_2^{'}(\theta_0) = 7/12 - [\beta_2(\theta_0) - 7/12]$ takes (e.g.) the $\theta_0 = 119^\circ$ near-FCC lattice  [Fig.\ \ref{fig:phimax}(b)] into a near-HCP lattice.

\section{Jammed Packings}
\label{sec:jammed}

\subsection{Molecular dynamics simulation method}

We now examine how our model trimers solidify under dynamic compression using molecular dynamics (MD) simulations.
Each simulated trimer contains three monomers of mass $m$.  
Trimers' bond lengths and angles are held fixed by holonomic constraints.
Monomers on different trimers interact via a harmonic potential $U_{H}(r) = 5u_0 (1 - r/\sigma)^2 \Theta(\sigma-r)$, where $u_0$ is the energy scale of the pair interactions, $\sigma$ is monomer diameter, and $\Theta$ is the Heaviside step function.

Initial states are generated by placing $n_{tri} = 1333$ trimers randomly within a cubic cell at a packing fraction $\phi_0 = \exp(-2/3)\phi_{cp}$.
Periodic boundary conditions applied along all three directions and Newton's equations of motion are integrated with a timestep $\delta t = .005\tau$, where the unit of time is $\tau=\sqrt{m\sigma^2/u_0}$.
Systems are equlibrated at finite temperature until intertrimer structure has converged, then rapidly cooled to $T=0$.
Then they are hydrostatically compressed at a true strain rate $\dot{\epsilon}$, i.e.\ the cell side length $L$ is varied as $L = L_0 \exp(-\dot\epsilon t)$.
To maintain near-zero temperature during compression, we employ overdamped dynamics with the equation of motion
\begin{equation}
m\ddot{\vec{r_i}} = \vec{F} - \gamma \dot{\vec{r_i}} +h(\{ \vec{r} , \dot\vec{r} \})
\label{eq:eom}
\end{equation}
where $\vec{r}_i$ is the position of monomer $i$, $\vec{F}$ is the force arising from the harmonic pair interactions, the damping coefficient $\gamma = 10^4\dot\epsilon$, and the $h(\{ \vec{r} , \dot\vec{r} \})$ term enforces trimer rigidity \cite{kamberaj05}.
To access the quasistatic limit, we also perform compression runs wherein compression is halted at equal increments of $\ln(\phi/\phi_0)$ [i.e.\ equal volumetric strain intervals] and followed by energy minimization as is standard in studies of jamming \cite{ohern02}.
As in our recent study of 2D bent-core trimers \cite{griffith18}, jamming is defined to occur when the nonkinetic part of the pressure $P$ exceeds $P_{thres}= 10^{-4}u_0/\sigma^2$ \cite{phiJdef}.
All MD simulations are performed using LAMMPS \cite{plimpton95}.

\subsection{Variation of $\phi_J$ with $\theta_0$ and strain rate}

\begin{figure}[htbp]
\includegraphics[width=3.25in]{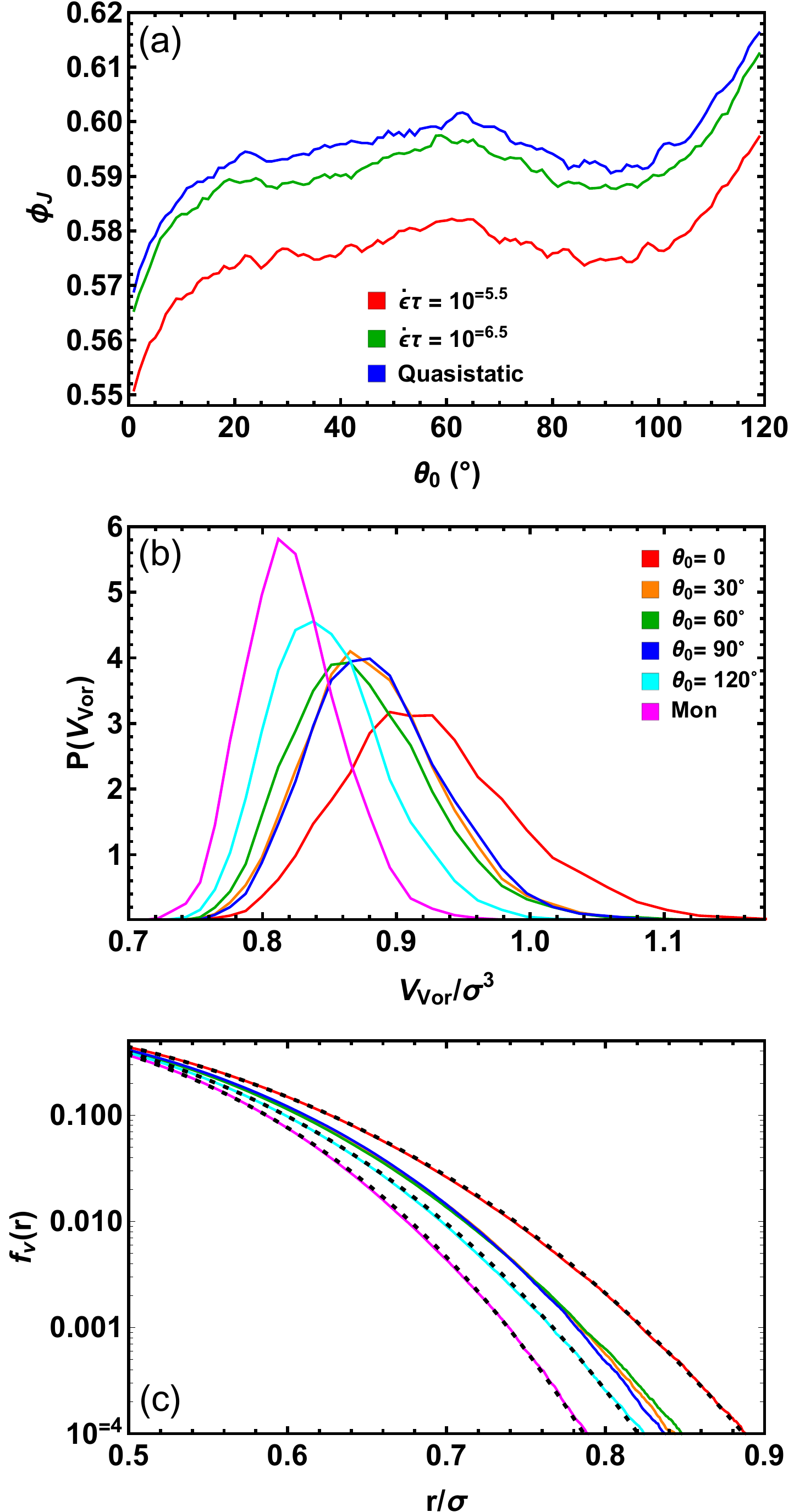}
\caption{Effect of the bond angle and compression rate on jamming of bent-core tangent-sphere trimers.  Panel (a) shows results for $\phi_{J}(\theta_0; \dot{\epsilon})$, while panels (b) and (c) respectively show results for $P(V_{\rm Vor})$ and $f_v(r)$ in selected quasistatically compressed, marginally jammed systems.  In panels (b-c), the magenta curves show results for monomers, and the dotted black curves show fits to Eq.\ \ref{eq:fvform}.}
\label{fig:phiJ}
\end{figure}

Figure \ref{fig:phiJ}(a) shows $\phi_J(\theta_0)$ for three different compression protocols.
The compression-rate-dependence is typical for granular materials \cite{ciamarra10}.
The variation in $\partial \phi_J(\theta_0; \dot\epsilon)/\partial\dot\epsilon$ with $\theta_0$ is small compared to the statistical noise in $\phi_J$, suggesting that the coupling of compression-rate- and particle-shape-driven effects is weak.
We will focus on results for quasistatically compressed systems for the remainder of this paper.

Three key results are immediately apparent from the variation of $\phi_J$ with $\theta_0$.
First, all trimers have $\phi_J$ significantly below the monomeric value ($\phi_J^{mon} \simeq  .637$ \cite{ohern02}).
This indicates that the quenched bond-length and bond-angle constraints always strongly promote jamming.
Second, $\partial\phi_J/\partial \theta_0$ is large and positive for $\theta_0 \lesssim 20^\circ$ and for  $\theta_0 \gtrsim 100^\circ$, but much smaller in the range $20^\circ \lesssim \theta_0 \lesssim 100^\circ$.
Third, $\phi_J(\theta_0)$ does not track $\phi_{max}(\theta_0)$; instead, its variation appears to arise from shape-dependent jamming mechanisms.
Explaining these results will be our focus for the remainder of this section.

The significantly lower $\phi_J$ for trimers with smaller $\theta_0$ strongly suggests that their marginally jammed (MJ) states are more disordered.
One simple metric for systems' degree of disorder is the breadth of their Voronoi volume distributions.
Fig.\ \ref{fig:phiJ}(b) shows the Voronoi-volume probability distributions $P(V_{\rm Vor})$ for selected $\theta_0$ (calculated using voro++ \cite{voro}) and contrasts them to those for a MJ system of $3n_{tri}$ monodisperse monomers; note that the latter jam at $\phi_J^{mon}$ for our quasistatic compression protocol.
Systems with lower $\phi_J$ are more disordered in the sense that their $P(V_{\rm Vor})$ distributions are broader and have longer tails.
However, the distributions even for the systems with the highest $\phi_J$ remain typical of those found in amorphous solids \cite{kumar05}, suggesting that none of the systems exhibit significant crystallization.
Indeed all results are consistent with trends found in monomeric hard- and soft-sphere systems \cite{torquato00,ciamarra10}; the distinction is that in our case the variations in $\phi_J$ and  $P(V_{\rm Vor})$ result from varying $\theta_0$ instead of from varying systems' preparation protocol.

Another way to interrogate the $\theta_0$- and rate-dependence of $\phi_J$ is to examine the statistical properties of the voids within 
jammed systems.
Void size distributions have recently been shown to be intimately connected to to both the local packing geometries within and the uniformity of MJ states \cite{zachary11}.
Moreover, results for ellipsoids, rods, and semiflexible polymers \cite{donev04b,kyrylyuk11,hoy17} suggest that the larger aspect ratio of small-$\theta_0$ trimers will lead to jammed states with an excess of large voids.
We calculated the fraction $f_{\rm v}(r)$ of empty space lying a distance $d > r$ away from the interior of any monomer  by sampling a large number of randomly placed points within our MJ configurations.
By definition, $f_{\rm v}(.5\sigma) = 1 - \phi$, and for $r > .5\sigma$ $f_{\rm v}$ declines monotonically with increasing $r$.
Fig.\ \ref{fig:phiJ}(c) shows $f_{\rm v}(r)$ for the same systems analyzed in Fig.\ \ref{fig:phiJ}(b).
Data for all systems are remarkably well fit by the simple 3-parameter functional form
\begin{equation}
f_{\rm v}(r) = A \exp\left[ -\left( \displaystyle\frac{r}{r_0} \right)^b  \right],
\label{eq:fvform}
\end{equation}
where $r_0$ is slightly less than $.5\sigma$ and $A \simeq (1 - \phi_J)\exp\left[(.5\sigma/r_0)^b\right]$.
We find that the exponent $b$ increases from $\sim 15/4$ to $\sim 9/2$ with increasing $\phi_J$ (e.g.\ with increasing $\theta_0$ for $\theta_0 \lesssim 20^\circ$ and for  $\theta_0 \gtrsim 100^\circ$).
However, these differences in MJ systems' void structure are seemingly quantitative rather than qualitative.

\begin{figure}[h!]
\includegraphics[width=3.375in]{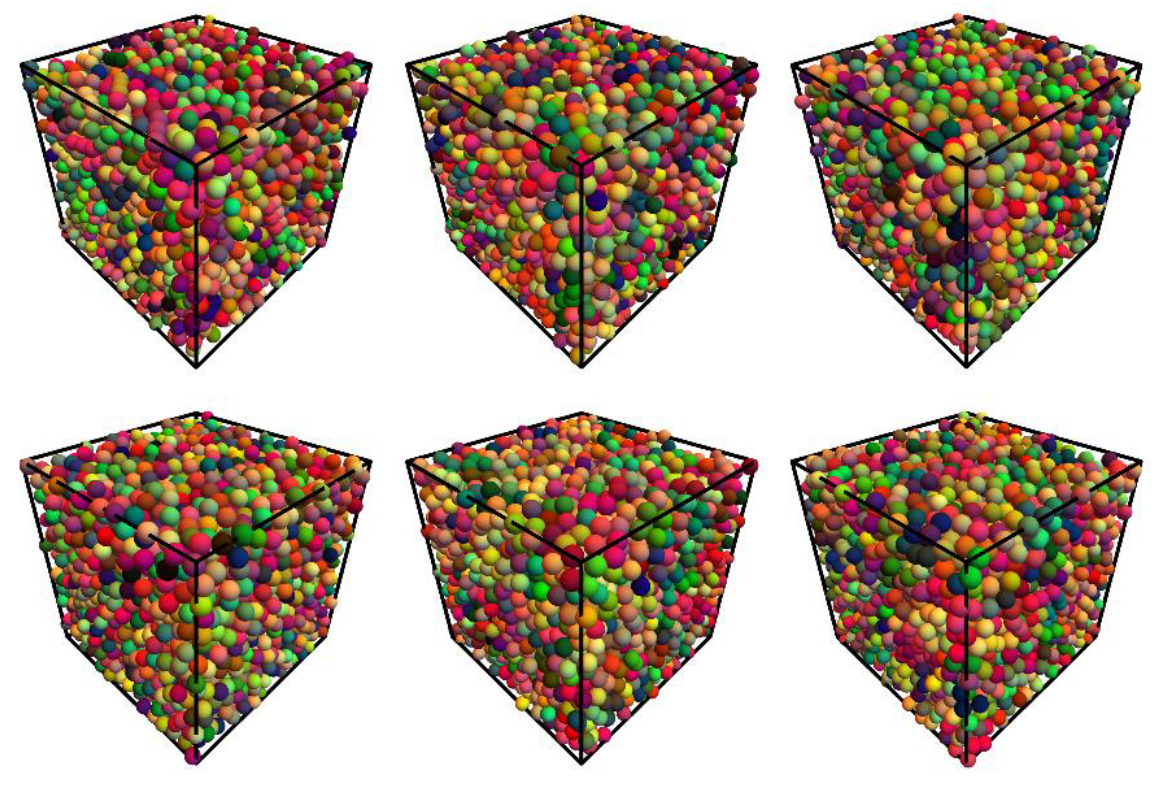}
\caption{Marginally jammed states for (clockwise from upper left) $\theta_0 = 0,\ \rm 30^\circ,\ 60^\circ,\ 90^\circ,\ \rm{and}\ 120^\circ$ trimers, and monomers.  Different colors indicate different trimers, but otherwise the color of each trimer is chosen randomly.}
\label{fig:margpic}
\end{figure}

Taken together, the above results imply that bent-core trimers are a model system in which both $\phi_J$ and the degree of MJ states' disorder can be tuned by varying $\theta_0$.  
However, it is less clear how/why these effects arise.
Visual inspection of MJ states (Fig.\ \ref{fig:margpic}) shows no obvious structural differences between systems with different $\theta_0$ or between trimeric and monomeric systems.

\subsection{Ordering within marginally jammed states}

\begin{figure}[h!]
\includegraphics[width=2.94in]{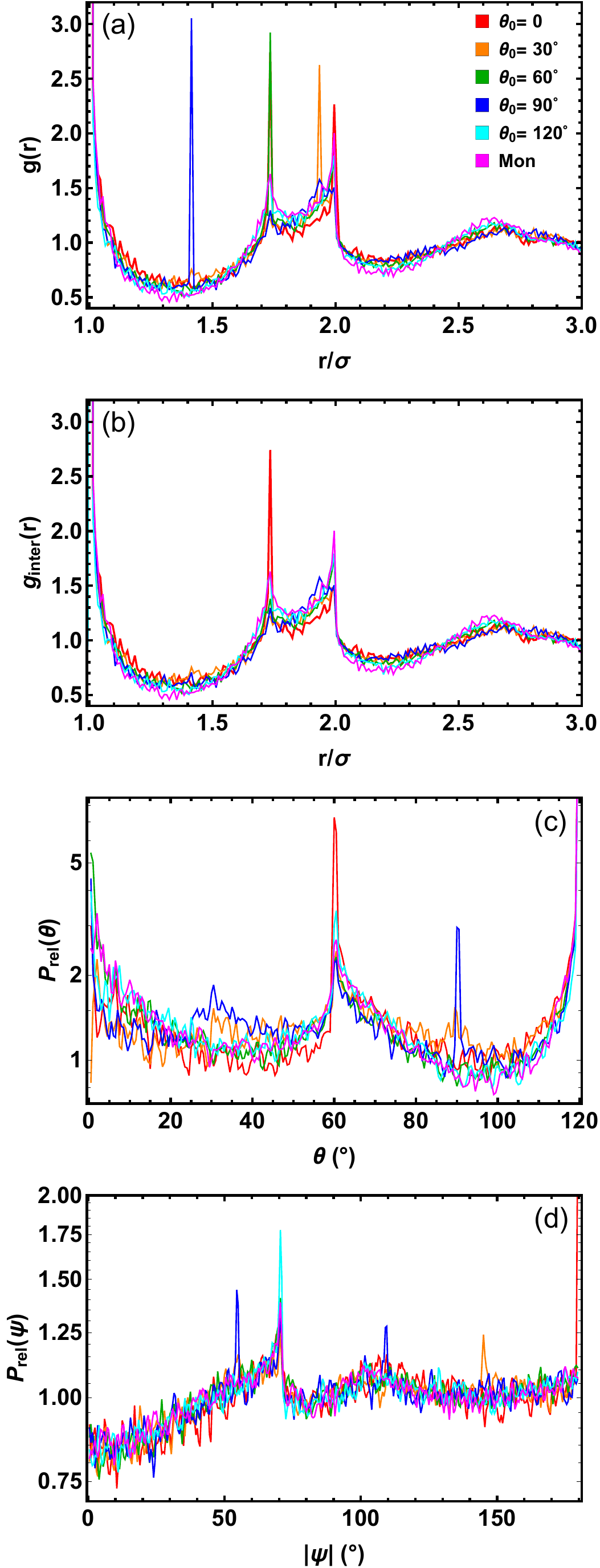}
\caption{Local structure of marginally jammed states.  Panels (a-b) show the pair correlation function $g(r)$ and its intermolecular contribution $g_{inter}(r)$.  Panels (c-d) show the probability distributions $P(\theta)$ and $P(\Psi)$ for the contact and dihedral angles formed by $i-j-k$ triplets and $i-j-k-l$ quadruplets.  Both $P(\theta)$ and $P(\Psi)$ are normalized so that $P = 1$ for randomly configured systems.}
\label{fig:234body}
\end{figure}

To better isolate the origin of the trends shown in Fig.\ \ref{fig:phiJ}, we will examine monomer-monomer positional correlations at the 2-, 3-, and 4-body levels.
Figure \ref{fig:234body} illustrates the character of these correlations in selected systems' MJ states.
Panel (a) shows the total pair correlation function $g(r)$.  
The most striking feature is the sharp peaks at $r = d_{13}(R, \theta_0)$, where $d_{13}(R, \theta_0) = 2R\cos(\theta_0/2)\sigma$ is the distance between a trimer's end monomers.
Another noteworthy feature is that $g(r)$ is nearly independent of $\theta_0$ for $r > 2\sigma$ despite the systems' very different densities.
The slight differences for $r > 2\sigma$ -- more prominent maxima and minima for systems with larger $\phi_J$ -- are consistent with these systems' slightly greater order.

Panel (b) shows the intermolecular contribution to the pair correlation function [$g_{inter}(r)$].
The absence of clear peaks at $r = \sqrt{2}\sigma$ and $r = \sqrt{8/3}\sigma$ (which are respectively the second-nearest-neighbor distances in FCC and HCP lattices) for all $\theta_0$ indicates that even locally close-packed order is strongly suppressed in these systems. 
Surprisingly, $\theta_0 = 0$ systems (which have the lowest $\phi_J$ and broadest distribution of void and Voronoi volumes) also have the sharpest peak in $g_{inter}(r)$ of any of our representative systems; this peak occurs at $r \simeq \sqrt{3}\sigma$ and will be discussed further below.
Otherwise, the most prominent $\theta_0$-dependent differences are for $1.2 \lesssim r/\sigma \lesssim 1.6$.  
Systems with lower $\phi_J$ have larger $g_{inter}(r)$ in this range, indicating that the distinction between monomers' first and second coordination shells is sharper for systems with  higher $\phi_J$.

Panel (c) shows the probability distribution $P(\theta)$ for the angles
\begin{equation}
\theta = \cos^{-1}\left( \displaystyle\frac{ \vec{b}_{ij} \cdot \vec{b}_{jk} }{b_{ij} b_{jk} }  \right)
\label{eq:psi}
\end{equation}
formed by contacting $i-j-k$ triplets that do not all belong to the same trimer \cite{foottri}.
Here $\vec{b}_{ij} = \vec{r}_j - \vec{r}_i$ and $\vec{b}_{jk} = \vec{r}_k - \vec{r}_j$, where ($\vec{r}_i,\ \vec{r}_j, \vec{r}_k$)  are the positions of monomers $i,\ j,\ \rm{and}\ k$, and the pairs [$(i, j),\ \textrm{and}\ (j, k)$] are each in contact, i.e.\ $b_{ij}\ \textrm{and}\ b_{jk}$ are each $\leq \sigma$.
The sharp peaks in $P(\theta_0)$ at $\theta_0 \simeq 60^\circ\ \rm{and}\ 120^\circ$ correspond to small 3-sphere subunits of the triangular lattice, e.g.\ closed equilateral triangles have $\theta_0 = 120^\circ$.
Surprisingly, the peak at $\theta_0 = 60^\circ$ is sharpest for $\theta_0 = 0$ trimers; this feature will be 
associated with these systems' distinctive $g_{inter}(r)$ below.
Triplets forming three sides of a square are far less common.
The peak in $P(\theta)$ at $\theta = 90^\circ$ is prominent only for $\theta_0 = 90^\circ$ where trimers \textit{always} form three sides of a square and another monomer (which belongs to an intertrimer triplet and hence gets counted in $P(\theta)$ \cite{foottri}) often completes it.

Panel (d) shows the probability distribution $P(\psi)$ for the dihedral angles
\footnotesize
\begin{equation}
\psi = \rm{atan2}\left[ \left( (\vec{b}_{ij} \times \vec{b}_{jk} )\times (\vec{b}_{jk} \times \vec{b}_{kl} ) \right) \cdot \hat{b}_{kl} \ , \ (\vec{b}_{ij} \times \vec{b}_{jk} ) \cdot (\vec{b}_{jk} \times \vec{b}_{kl} ) \right]
\label{eq:psi}
\end{equation}
\normalsize
formed by contacting $i-j-k-l$ quadruplets \cite{dihed}.  
Here $\vec{b}_{ij}$ and $\vec{b}_{jk}$ are as defined above, and $\vec{b}_{kl} = \vec{r}_l - \vec{r}_k$, where $\vec{r}_l$ is the position of monomer $l$.  
The pairs [$(i, j),\ (j, k),\ \textrm{and}\ (k,l)$] must be in contact, i.e.\ $b_{ij},\ b_{jk},\ \textrm{and}\ b_{kl}$ must each be $\leq \sigma$.
The peaks at $|\Psi| = 54.7^\circ,\ 70.5^\circ,\ \textrm{and}\ 109.5^\circ$ have been previously observed in MJ configurations of monodisperse hard spheres \cite{anikeenko07} and flexible tangent-sphere polymers \cite{karayiannis09b};\ they indicate locally tetrahedral and/or polytetrahedral ordering.
$\theta_0 = 0$ trimers have a sharp peak at $|\Psi| = 180^\circ$ that the other systems lack.
This peak corresponds to planar \textit{trans} conformations \cite{dihed} and indicates that straight trimers have a tendency to form locally planar structures even though they remain disordered.
Other noteworthy $\theta_0$-dependent differences are that the peaks corresponding to tetrahedral quadruplets are sharpest for $\theta_0 = 90^\circ\ \rm{and}\ 120^\circ$.
The peak at $\theta_0 = \cos^{-1}(1/3) \simeq 70.5^\circ$ is particularly sharp for $\theta_0 = 120^\circ$ because these trimers automatically form three-fourths of an ideal tetrahedron; this commensurability with polytetrahedral order may be one reason why $\theta_0 = 120^\circ$ maximizes $\phi_J$.

\begin{figure}[h]
\includegraphics[width=3.25in]{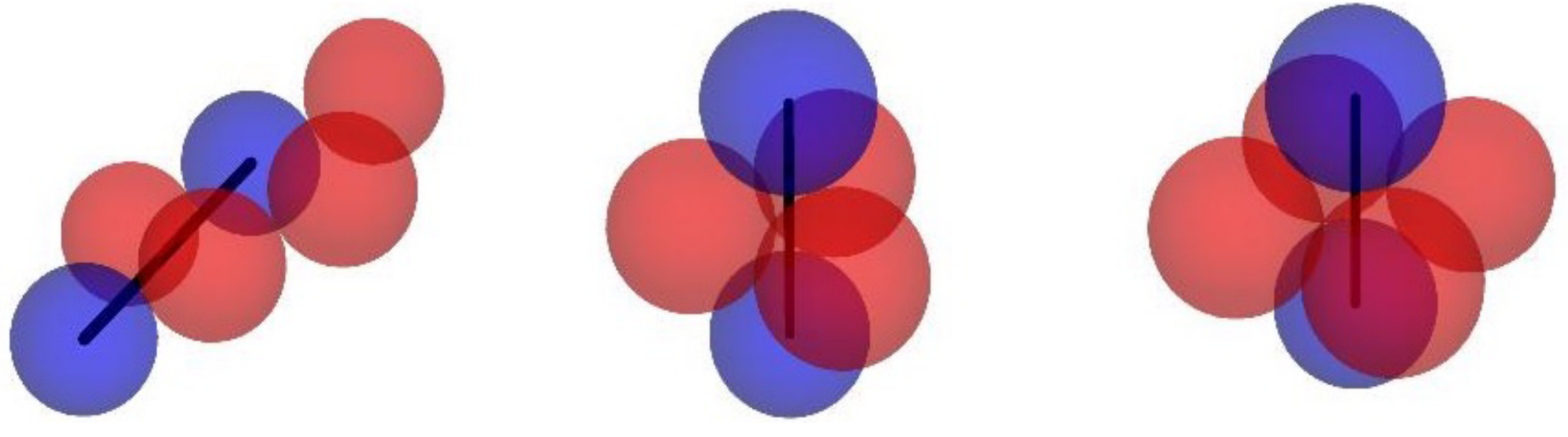}
\caption{Favored and disfavored close-packed motifs in marginally jammed states of bent-core trimers:\ aligned straight triplets, triangular bipyramid, octahedron.}
\label{fig:favdisfav}
\end{figure}

Taken together, the results in Fig.\ \ref{fig:234body} identify three motifs present in close-packed crystals that are particularly relevant to bent-core trimers' MJ states.
These motifs are shown in Figure \ref{fig:favdisfav}.
The leftmost is formed by two aligned straight triplets that occupy a small subset of a 2D triangular lattice;  the distance between the blue monomers is $\sqrt{3}\sigma$.
This motif is particularly prevalent in $\theta_0 = 0$ systems because straight trimers can easily form the abovementioned aligned triplets; its prevalence produces the strong peaks in $g_{inter}(r)$ at $r = \sqrt{3}\sigma$, in $P(\theta)$ at $\theta = 60^\circ$, and in $P(\Psi)$ at $|\Psi| = 180^\circ$.
The middle motif, known as the triangular bipyramid, is typical in HCP crystals; the distance between the blue monomers is $\sqrt{8/3}\sigma \simeq 1.63\sigma$.
Surprisingly, this motif is strongly suppressed in MJ states for all $\theta_0$ despite the fact that the two blue monomers can form a triplet [with $\theta = \cos^{-1}(1/3)$] with any of the red ones.
The rightmost motif, the octahedron, is formed by pairs of 2nd-nearest neighbors in a FCC lattice together with the four spheres that contact them both; the distance between the blue monomers is $\sqrt{2}\sigma$.
Given that the triangular bipyramid is not found in FCC lattices and the octahedron is not found in HCP lattices, these motifs are presumably suppressed by the same competition between FCC and HCP ordering that inhibits crystallization in athermal compression of monomers \cite{lubachevsky91}.

\subsection{Effect of concavity}
\label{subsec:concav}

Concavity is well known to promote jamming by increasing particle interlocking -- the influence of a monomer that contacts two bonded monomers belonging to a different grain increases with increasing $R$ \cite{schreck09}.
To examine the effect of concavity on trimer jamming, we contrast our above results for tangent-sphere trimers to those for overlapping-sphere trimers with $R < 1$.
Each trimer's excluded volume is
\footnotesize
\begin{equation}
v_{tri}(R,\theta_0) = \bigg{ \{ } \begin{array}{lcc}
3v_{mon} - 2v_{l}(R) & , & \theta_0 < \theta_{ov}(R) \\
\\
3v_{mon} - 2v_{l}(R) - v_{l}[d_{13}(R, \theta_0)] & , & \theta_0 \geq \theta_{ov}(R)
\end{array},
\end{equation}
\normalsize
where the volume of each monomer is $v_{mon} = \pi\sigma^3/6$, the volume of the lenses formed by overlapping spheres separated by a distance $r < \sigma$ is $v_{l}(r) = (\pi \sigma^3/12)(1-r/\sigma)^2(2+r/\sigma)$, and trimers' end monomers overlap only for  $\theta_0 \geq \theta_{ov}(R) \equiv 2\cos^{-1}[1/(2R)]$.
Here we study systems with $R = 0.856$, which matches the ratio of bond length to bead diameter in the widely used Kremer-Grest bead-spring polymer model \cite{kremer90}, is known to strongly inhibit crystallization, and has $\theta_{ov} = 108.526^\circ$.

\begin{figure}[htbp]
\includegraphics[width=3.25in]{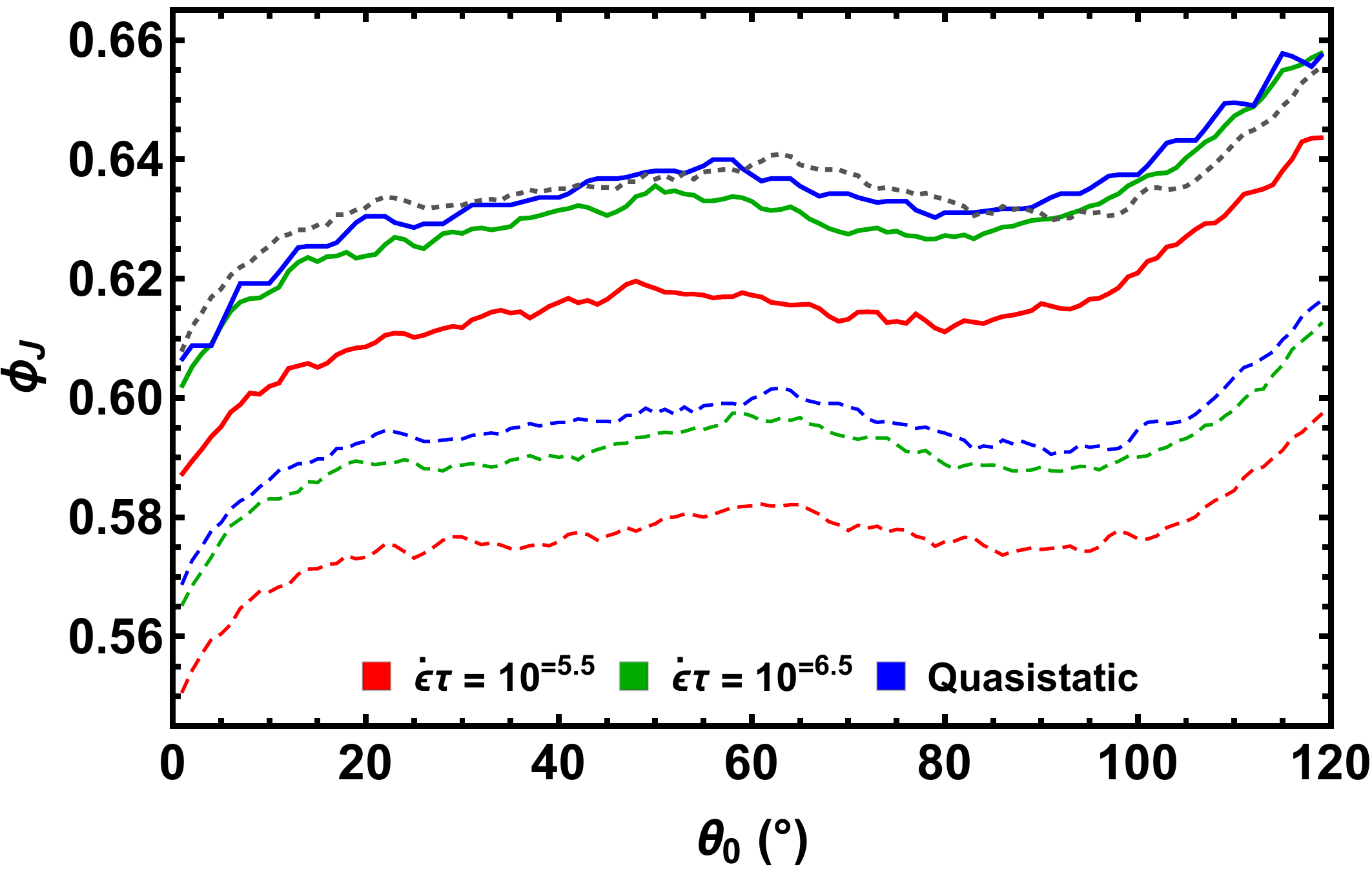}
\caption{Effect of monomer overlap on rate-dependent $\phi_J(\theta_0)$.  Solid curves indicate results for $R = 0.856$ trimers, dashed curves indicate the same $R = 1$ results shown in Fig.\ \ref{fig:phiJ}(a), and the gray dotted curve indicates the best fit of the quasistatic $R = 1$ results to Eq.\ \ref{eq:phiJR}: $s(.856) = .0392$.}
\label{fig:conc}
\end{figure}

Results for the rate-dependent $\phi_J(\theta_0, R)$ are shown in Figure \ref{fig:conc}.
Two principal effects of increasing monomer overlap (decreasing concavity) are immediately apparent.
First, the $\phi_J$ curves flatten as $R$ decreases, as they must.
Second, when $1-R \ll 1$ (as is the case for $R=0.856$), the effect of increasing overlap is -- to a first approximation -- a uniform upward shift of the $\phi_J$ curves, i.e.\
\begin{equation}
\phi_J^R(\theta_0) = \phi_J^1(\theta_0) + s(R)
\label{eq:phiJR}
\end{equation}
where $s$ increases with decreasing $R < 1$ but does not depend on $\theta_0$.  
The effectiveness of Eq.\ \ref{eq:phiJR} in describing the $\phi_J(\theta_0, R)$ data in Fig.\ \ref{fig:conc} suggests that  the decrease in bent-core trimers' $\phi_J$ with increasing $R$ is driven primarily by the increasing strength of monomer-dimer interlocking rather than by any structural features that depend directly on $\theta_0$.

\subsection{Comparison to results from 2D systems}

\begin{figure}[h]
\includegraphics[width=3.25in]{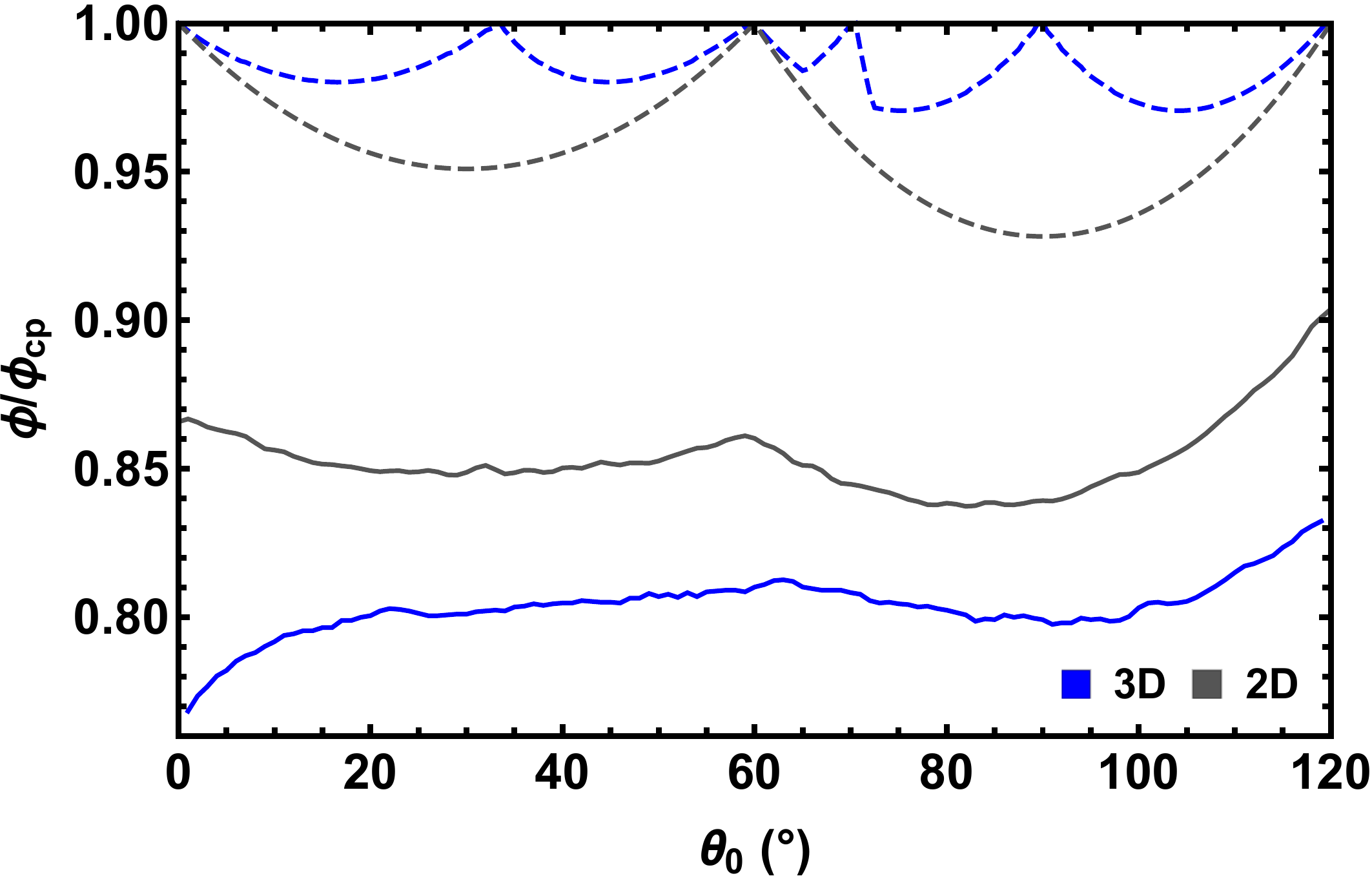}
\caption{Effect of spatial dimension on dense-packing and jamming of tangent-sphere trimers.  Solid curves show $\phi_J(\theta_0)$ for quasistatic compression while dashed curves show $\phi_{max}(\theta_0)$.  The 2D results are from Ref.\ \cite{griffith18}.  Packing fractions are scaled by their values for close-packed crystals [$\pi/(2\sqrt{3})$ in 2D, $\pi/\sqrt{18}$ in 3D].}  
\label{fig:dim}
\end{figure}

Solidification of particulate matter under athermal compression is well known to exhibit a striking dependence on spatial dimension.
Monodisperse disks readily crystallize whereas monodisperse spheres typically form disordered jammed states \cite{lubachevsky91,torquato00}.
Here we discuss how spatial dimension affects athermal bent-core-trimer solidification.
Figure \ref{fig:dim} contrasts $\phi_J(\theta_0)$ and $\phi_{max}(\theta_0)$ for 2D and 3D trimers.
Three key results are apparent.
First, the quenched bond-length and bond-angle constraints produce comparable enhancement of jamming.
The values of $\phi_J(\theta_0)/\phi_{cp}$ averaged over all $\theta_0$ are $.853$ in 2D and $.803$ in 3D.
The difference between these fractions in comparable to the difference between the values of  $\phi_J^{mon}/\phi_{cp}$ for amorphous monomeric MJ states ($\phi_J^{mon}/\phi_{cp} \simeq .92$ in 2D, $\simeq .86$ in 3D \cite{ohern02}).
Second, in 2D trends in $\phi_J(\theta_0)$ track those in $\phi_{max}(\theta_0)$ throughout the full range $0 \leq \theta_0 \leq 120^\circ$, while in 3D they do not.
In particular, 3D straight trimers ($\theta_0 = 0$) \textit{can} close-pack yet produce the global minimum in $\phi_J(\theta_0)$.
This differerence arises partially from differences in crystallizability; 2D $\theta_0 = 0$ trimers form MJ states possessing a moderate degree of crystalline order \cite{griffith18}, whereas in 3D their MJ states are maximally disordered as discussed above.
Third, in both 2D and 3D compactness and symmetry promote dense packing, i.e.\ $\phi_J$ is maximized for $\theta_0 = 120^\circ$ trimers since they are subunits of the triangular lattice.

Numerous previous studies of the jamming of anisotropic grains have found that $\phi_J$ decreases with increasing grain aspect ratio $\mathcal{A}$.
Others have found nonmonotonic behavior; for grains such as ellipses and dimers, $\phi_J$ increases with increasing $\mathcal{A}$ for small $\mathcal{A}$, passes through a single maximum at some $\mathcal{A} = \mathcal{A}^*$, then decreases with further increasing $\mathcal{A}$ \cite{donev04b,mailman09,borsonyi13}.
The results in Figs.\ \ref{fig:conc}-\ref{fig:dim} indicate that no comparably simple explanation of the variation of $\phi_J(\theta_0)$ in terms of $\mathcal{A}$ is possible for bent-core trimers [which have $\mathcal{A}(R, \theta_0) = 2R\cos(\theta_0/2) + 1$].
While their $|\partial A/\partial \theta_0| \to 0$ as $\theta_0 \to 0$, their $|\partial \phi_J(\theta_0)/\partial \theta_0|$ is maximized as $\theta_0 \to 0$.
A more likely explanation of 3D bent-core trimers' decrease in $\phi_J(\theta_0)$ as $\theta \to 0$ is the decrease in their \textit{effective} configurational freedom as they approach linearity.
Specifically, in 3D the middle monomer in a bent trimer can relax away from obstacles by rotating about the line connecting the end monomers (even if the end monomers are held fixed), whereas the middle monomer in a straight trimer cannot.
In 2D, no such rotational-relaxation mechanisms are available.
Note that similar decreases in effective configurational freedom as ``molecules'' straighten have previously been associated with decreasing $\phi_J$ and increasing $T_g$ in model polymeric systems \cite{hoy17,plaza17}.

\section{Discussion and Conclusions}
\label{sec:discuss}

In this paper we showed that simply structured bent-core tangent-sphere trimer crystals (with bases of $n_{tri} \leq 2$ trimers) have maximum packing fractions above $0.97\phi_{cp}$ for all $\theta_0$.
While excluding the possibility that denser crystals can be found by considering lattices with larger bases would require a proof by numerical exhaustion like that of Hales \cite{hales98}, previous results for a wide variety of particle shapes \cite{kuperberg90,torquato12} suggest that $n_{tri} = 2$ is sufficiently large to identify the crystals reported here as (putatively) maximally dense.

We showed that incommensurability of $\theta_0$ with 3D close-packing does not by itself frustrate crystallization.
Instead, trimers are able to arrange into periodic structures composed of close-packed  bilayers or trilayers of triangular-lattice planes, separated by ``gap layers'' that accomodate the incommensurability.
Except for the narrow range $70.5288^\circ \leq \theta_0 \leq 72.4530^\circ$, such stackings are maximally dense.
In this narrow range, the incommensurability is instead accommodated by deviation of the stacked planes away from triangular-lattice order.

Because obtaining crystalline bent-core-trimer systems in experiments may be challenging, we contrasted their dense crystalline packings to the marginally-jammed packings they form under athermal compression.
Our results make it apparent that two distinct sets of factors act in concert to promote jamming in these systems:\ \textbf{(i)} the same factors that promote jamming in monodisperse hard-sphere systems (e.g.\ the FCC-HCP competition \cite{lubachevsky91}), and \textbf{(ii)} other $\theta_0$-dependent factors associated with the quenched 2- and 3-body constraints inherent to bent-core trimer structure (Fig.\ \ref{fig:trimermodel}).  
That set \textbf{(ii)} always further promotes jamming [i.e.\ $\phi_J(\theta_0) < \phi_J^{mon}$ for all $\theta_0$] is surprising:\ although the quenched constraints reduce the dimensionality of a system's configuration space, a larger fraction of that space corresponds to crystalline order if $\theta_0$ is commensurable with close-packing.
Other factors, such as the decrease in trimers' effective configurational freedom as they straighten, are seemingly more important than commensurability.

The dependence of granular materials' macroscopic properties on the shape of the grains composing them has attracted great interest over the past decade \cite{damasceno12,torquato12,cersonsky18,harrington18,borsonyi13,murphy19}.
The results presented in this paper suggest that bent-core trimers are a model granular system in which both $\phi_J$ and the local ordering of jammed states can be tuned by varying a single particle-shape parameter ($\theta_0$).
Moreover, they should be relatively easy to synthesize using readily available techniques \cite{scalfani14,harrington18,olson02}.
Future work will examine how jammed bent-core trimer systems' mechanical and acoustic properties depend on $\theta_0$.

This material is based upon work supported by the National Science Foundation under Grant DMR-1555242.


\begin{thebibliography}{53}
\bibitem [{fam()}]{famIVnote0}%
  \BibitemOpen
  \href@noop {} {}\bibinfo {note} {In the later stages of our analyses, we used
  a finer $\alpha$-increment and a restricted $\beta$-range to fully
  characterize family IV; see Section \ref{subsec:dtc}.}\BibitemShut {Stop}%
\bibitem [{phi()}]{phiJdef}%
  \BibitemOpen
  \href@noop {} {}\bibinfo {note} {Choosing a lower (higher) value of
  $P_{thres}$ lowers (raises) $\phi_J(\theta_0)$, but does not qualitatively
  change any of the results presented herein. We choose to identify jamming
  with the emergence of a finite bulk modulus rather than with the vanishing of
  soft modes because proper handling of soft modes associated with trimeric
  ``flippers'' (configurations in which the end monomers are mechanically
  stabilized but the middle one remains free to rotate about the line
  connecting their centroids \cite{karayiannis08}) is nontrivial.}\BibitemShut
  {Stop}%
\bibitem [{foo()}]{foottri}%
  \BibitemOpen
  \href@noop {} {}\bibinfo {note} {Triplets belonging to the same trimer are
  excluded since they would just make trivial $\delta$-function contributions
  to $P(\theta)$ at $\theta = \theta_0$.}\BibitemShut {Stop}%
\end{thebibliography}

%

\end{document}